\documentclass[11pt]{article}
\usepackage{jheppub}

\input{epsf}
\usepackage{epsfig}
\usepackage{amssymb}
\usepackage{amsfonts}
\usepackage{amsbsy}
\usepackage[all]{xy}
\usepackage{amsmath}
\usepackage{dsfont}
\usepackage{bbm}
\usepackage{fixmath}
\usepackage{upgreek}
\usepackage{amssymb,amscd}
\usepackage{mathrsfs}
\usepackage{amsmath,amsthm}
\usepackage{MnSymbol}
\usepackage{slashed}
\usepackage{dsfont}
\usepackage{tikz}
\usetikzlibrary{positioning}
\usetikzlibrary{shapes.geometric, arrows}
\tikzstyle{rect} = [rectangle,rounded corners,minimum width=3cm, minimum height=1cm, text centered, draw=black]
\tikzstyle{arrow} = [thick,->,>=stealth]
\usepackage{stackrel}

\tikzstyle{hex} = [regular polygon,regular polygon sides=6, draw,
    text width=2em, text centered]
\tikzstyle{hexs}= [regular polygon,regular polygon sides=5, draw,
    text width=1.7em, text centered]
\tikzstyle{line} = [draw, -latex']
\def\be{ \begin{equation} }
\def\ee{ \end{equation}}

\def\log{{\rm log}}

\def\half{\frac{1}{2}}

\def\fBPS{\overline{\underline{\Omega}}}

\def\one{{\hbox{ 1\kern-.8mm l}}}

\def\vx{{\vec{x}}}
\def\vp{{\vec{p}}}
\def\vq{{\vec{q}}}

\def\vk{{\vec{k}}}
\def\vw{{\vec{w}}}

\def\vv{{\text{v}}}

\def\CA{{\cal A}}
\def\CB{{\cal B}}

\def\CD {{\cal D}}
\def\CE {{\cal E}}

\def\CH {{\cal H}}
\def\CI {{\cal I}}
\def\CJ {{\cal J}}
\def\CK {{\cal K}}
\def\CL {{\cal L}}
\def\CM {{\cal M}}
\def\CN {{\cal N}}

\def\CP {{\cal P}}
\def\CR {{\cal R}}
\def\CV {{\cal V}}
\def\CW {{\cal W}}
\def\CY {{\cal X}}

\def\CE {{\cal E}}

\def\CH {{\cal H}}
\def\CI {{{\cal I}}}
\def\CB {{\cal B}}
\def\CQ {{\cal Q}}
\def\CS {{\cal S}}

\def\CX {{\cal X}}
\def\CY {{\cal Y}}

\def\IC{\mathbb{C}}

\def\IJ{\mathbb{J}}

\def\IR{{\mathbb{R}}}

\def\IZ{{\mathbb{Z}}}

\def\fa{\mathfrak{a}}
\def\fb{\mathfrak{b}}

\def\ft{\mathfrak{t}}

\def\ft{\mathfrak{t}}

\def\fX{\mathfrak{X}}

\def\rmk#1{\bigskip\noindent{\bf Remark} }
\def\aside#1{\bigskip\noindent{\bf Aside} }

\newcommand\fro{{\overline{\underline{\Omega}}}}

\def\fMM{\overline{\underline{\mathcal{M}}}}

\title{Index-Like Theorems from Line Defect Vevs}

\author[a]{T. Daniel Brennan} 
\author[a]{Gregory W. Moore}
\affiliation[a]{NHETC and
Department of Physics and Astronomy, Rutgers University \\
126 Frelinghuysen Rd., Piscataway NJ 08855, USA}

\emailAdd{tdanielbrennan@physics.rutgers.edu}
\emailAdd{gwmoore@physics.rutgers.edu}

\abstract{In this paper we investigate the relation between complexified Fenchel-Nielsen coordinates and spectral network coordinates on Seiberg-Witten moduli space.
The main technique is the comparison of exact expressions for
the expectation value of 't Hooft defects in certain 4D $SU(2)$ $\CN=2$   gauge theories.  We derive an index-like theorem for a class of Dirac operators on singular monopole moduli spaces. Our expression determines the indices of Dirac operators on singular monopole moduli spaces  in terms of
characteristic numbers for vector bundles over certain Kronheimer-Nakajima quiver varieties. 
}

\preprint{}
	
\begin{document}

\maketitle

\section{Introduction and Technical Summary}

Exact results in quantum field theories with four-dimensional $\CN=2$ supersymmetry
can lead to nontrivial and interesting mathematical predictions. In this paper
we consider exact results for line defect expectation values in theories which
are of class $\CS$ but also possess a Lagrangian formulation.

In general, class $\CS$ theories are
closely related to Hitchin systems and in these theories the Hitchin moduli
space $\CM$ is interpreted as a space of vacua when the 4D theory is compactified
on a circle. If a line defect $L$ preserves four supersymmetries and is wrapped on
this circle then the associated vev $\langle L \rangle$ is a holomorphic function
on $\CM$ in a complex structure determined by the supersymmetry preserved by $L$.
The preserved supersymmetry may be characterized by a phase $\zeta$, which may be viewed
as an element of the twistor sphere: $\zeta$ also determines a complex structure on
$\CM$. We will denote the space $\CM$ with complex structure determined by $\zeta $ as $\CM_\zeta$. In the theories under consideration the holomorphic function
$\langle L(\zeta) \rangle$ on $\CM_{\zeta}$ can be computed, exactly, in three
different ways. This paper explores some consequences of comparing the
resulting exact expressions.

For theories of class $\CS$ 
one of the exact methods expresses $\langle L(\zeta)\rangle$ 
in terms of ``spectral network coordinates'' on $\CM_\zeta$ \cite{Gaiotto:2010be,Gaiotto:2012db,Gaiotto:2012rg}. 
These coordinates are generalizations of well-known cluster, shear, and Fock-Goncharov
coordinates. They are functions on the twistor space and, restricted to a fiber $\CM_\zeta$, are holmorphic Darboux coordinates in complex structure $\zeta$. 
See Section \ref{sec:3} below for a description of ``spectral network coordinates.''
We refer to the exact result for  $\langle L(\zeta) \rangle$ in these coordinates as the
``Darboux expansion.'' See equation \eqref{firstexp} below for the detailed formula.

On the other hand, thanks to the AGT correspondence, $\langle L(\zeta) \rangle$ can also be
expressed in terms of complexified Fenchel-Nielsen coordinates \cite{Drukker:2009id,Alday:2009fs}. 
These are also holomorphic Darboux coordinates on $\CM_\zeta$. 
See Section \ref{sec:CFN} for a description of Fenchel-Nielsen coordinates.
Thus, by comparing answers for a finite set of line defects one can express one
coordinate system in terms of the other.  The transformation turns out to be rather
nontrivial.

In theories of class $\CS$ that also have a Lagrangian description
one can also evaluate $\langle L(\zeta) \rangle$ exactly using localization techniques. These localization
techniques are only understood fully for certain line defects known as pure 't Hooft operators,
and are only valid in weak-coupling regimes in the Coulomb branch. \footnote{Mathematically by weak-coupling regions we refer to certain
asymptotic regimes of $\CM_\zeta$.} The function $\langle L(\zeta) \rangle$ for an 't Hooft line defect, when
evaluated using localization, turns out, again, to be expressed
naturally in terms of complexified Fenchel-Nielsen coordinates and moreover involves interesting
equivariant characteristic numbers of Kronheimer-Nakajima varieties.

At the same time, in the  weak coupling region of a class $\CS$ theory the coefficients of the Darboux expansion
have an interpretation in terms of indices of certain Dirac-like operators defined on moduli spaces
of (singular) magnetic monopoles. Thus, by first
comparing line defect vev's to determine the change of variables from spectral network to
Fenchel-Nielsen coordinates and then using the above facts, we find unusual expressions
for the $L^2$-index of certain Dirac operators in terms of characteristic numbers of
Kronheimer-Nakajima varieties. This comparison yields our ``index-like'' theorem.

We should stress how unusual our ``index-like'' theorem is. 
In standard index theorems the index of a Dirac operator on one 
manifold is expressed in terms of a characteristic number of the 
relevant bundle on the same manifold. In the present case the 
($L^2$-) index of a Dirac operator is expressed in terms of 
sums of characteristic numbers on a different (but related) 
manifold. (Alternatively, an equivariant characteristic number is expressed as a sum of indices 
of Dirac operators.) 
Hence the qualifier ``index-like theorem."  The physical 
idea here is essentially trivial, but the mathematical statement 
seems to be fairly nontrivial and would seem to be challenging to prove from 
first principles. 

The remainder of this paper implements the above idea
in the special case of 4D $G=SU(2)$ $\CN=2$ supersymmetric gauge theories with adjoint or fundamental matter.
In the remainder of this introduction we explain the idea in a bit more detail.

\subsection{Technical Summary}

In the case of 4D $G=SU(2)$ $\CN=2$ supersymmetric gauge theories, there are two complexified-Fenchel-Nielsen coordinates which we will denote $\fa,\fb$.
From general principles, the expectation value of the 't Hooft defects can be expressed in Fenchel-Nielsen  coordinates as a Fourier expansion in $\fb$. More precisely, this can be written as \cite{Ito:2011ea,Gomis:2011pf}
\be
\langle L_{p,0}(\zeta)\rangle=\sum_{v\in \IZ_+\,:\, v\leq p}\cosh(\vv,\fb) (F(\fa))^{v}Z_{mono}(\fa,m,\epsilon;P,\vv)~,
\ee
where ${\rm v=diag}(v,-v)$ and $P={\rm diag}(p,-p)$ where $v,p\in \IZ$. On the right hand side the $\zeta$-dependence is captured by the use of complexified Fenchel-Nielsen coordinates on $\CM_\zeta$. \footnote{Usually complexified Fenchel-Nielsen coordinates are introduced as holomorphic coordinates, depending on a cutting system, of the character variety $\mathfrak{X}=Hom(\pi_1(C),G_\IC)\big\slash conj.$ for some complex gauge group $G_\IC$. In our case, $\CM_\zeta$ is isomorphic to $\fX$ for all $\zeta\neq0,\infty$ as a complex manifold, but not canonically. Our Fenchel-Nielsen coordinates will therefore also be functions on the twistor space of $\CM$ (with the fibers above $\zeta=0,\infty$ removed). When restricted to a fiber $\CM_\zeta$, they are holomorphic Darboux coordinates. It is in this way that they become comparable to spectral network coordinates. 
}

Here the expectation value above is expressed as a sum over monopole bubbling configurations where  $\cosh(\vv,\fb)F(\fa)^v$ encodes the contribution of bulk fields and $Z_{mono}(\fa,m,\epsilon;P,\vv)$ describes the contribution from the SQM that arises on the 't Hooft defect from bubbling \cite{Brennan:2018yuj}. 
 See \cite{Brennan:2018yuj,Brennan:2018moe,Brennan:2018rcn} for more background and explanation of notation.

In the localization computation of $\langle L_{p,0}\rangle$, $Z_{mono}(P,\vv)$ is given by a characteristic number of a  certain resolved Kronheimer Nakajima space \footnote{There is an additional subtlety with 4D $\CN=2$ $SU(N)$ theories with N$_f=2N$. See footnote \ref{Nf2N}  for more details.}
\be
Z_{mono}(\fa,m,\epsilon;P,\vv)=\lim_{\xi\to0}\int_{\widetilde\CM_{KN}^\xi(P,\vv)}e^{\omega +\mu_T}\widehat{A}_T(T\widetilde\CM_{KN}^\xi)\cdot C_{T\times T_F}(\CV(\CR))~.
\ee
Here   $\widetilde\CM_{KN}^\xi(P,\vv)$ is a certain resolved Kronheimer-Nakajima space determined by the line defect charge $(P)$ and core magnetic charge ($\vv$), $e^{\omega+\mu_T}$ induces the $T$-equivariant volume form on $\widetilde\CM_{KN}^\xi(P,\vv)$, $\widehat{A}_T(T\widetilde\CM^\xi_{KN})$ is the $T$-equivariant $\widehat{A}$-genus  that describes the contribution from the $\CN=2$ vectormultiplet and $C_{T\times T_F}(\CV(\CR))$ is a characteristic class related to the matter hypermultiplets where $T$ is the Cartan torus of the conserved global symmetry group of flavor, $R$-, and global gauge transformations. 
The equivariant integral can then be evaluated as a contour integral in an algebraic torus whose poles are enumerated by Young tableaux \cite{Nekrasov:2002qd,Moore:1997dj,Moore:1998et}.  \footnote{
See Section \ref{sec:charnum} for exact definitions and more details.
}

On the other hand, using the class $\CS$ technology, the expectation value of a supersymmetric line defect can be computed by the trace of the holonomy of a flat $SL(N;\IC)$ connection along a corresponding curve $\gamma$ in an associated Riemann surface $C$.  Spectral networks express
 the expectation value of such 4D line defects as 
\be\label{firstexp}
\langle L_{p,0}\rangle_{u \in\CB}=\sum_{\gamma \in \Gamma} \fro(\gamma,L_{p,0};u)\CY_\gamma~,
\ee
where $\fro(\gamma,L_{p,0};u)$ are framed BPS indices, $\CY_\gamma$ are Darboux functions  on the moduli space of flat $SL(N;\IC)$ connections on $C$ associated to the physical charge $\gamma \in \Gamma$, and $\Gamma $ is a torsor of the IR charge lattice  \cite{Gaiotto:2010be}. 

In the semiclassical limit of the theories we are considering,
the framed BPS indices of 't Hooft defects can be identified with the index of a twisted Dirac operator on singular monopole moduli space \cite{Manton:1981mp,Brennan:2016znk,Moore:2015szp,Moore:2015qyu,Tong:2014yla,Gauntlett:1999vc,Gauntlett:2000ks,Manton:1993aa}. Locally on moduli space we can decompose $\gamma$ non-canonically into magnetic, electric, and flavor charge $\gamma=\gamma_m\oplus \gamma_e\oplus \gamma_f$. Then we have:
\be
\fro(\gamma,L_{p,0};u)={\rm Ind}_{L^2}\big[\slashed{D}^{\CY}\big]_{\CE_{\rm matter}\otimes S\fMM(P,\gamma_m,u)}^{\gamma_e\oplus\gamma_f}~.
\ee
Here the superscript $\gamma_e\oplus\gamma_f$ denotes the associated eigenspace of the $L^2$ index of   $\slashed{D}^{\CY}$, a Dirac operator modified by adding Clifford multiplication by a hyperholomophic vector field defined by $\CY\in \ft_\IC$. The Dirac operator acts on sections of $\CE_{\rm matter}\otimes S\fMM(P,\gamma_m,u)$ where $S\fMM(P,\gamma_m,u)$ is the spinor bundle on the singular monopole moduli space $\fMM(P,\gamma_m,u)$ and $\CE_{\rm matter}\to \fMM(P,\gamma_m,u)$ is a vector bundle over it related to hypermultiplet zero-modes. 
\footnote{See Section \ref{sec:SCFBPS} for exact definitions.} 


Thus, by comparing the expectation value of 't Hooft defects computed via localization and spectral network techniques in a weak coupling limit, we can derive a relation between characteristic numbers of Kronheimer-Nakajima spaces and indices of Dirac operators on singular monopole moduli space:
\begin{align}\begin{split}\label{GeneralForm}
&\sum_{\gamma \in \Gamma}
\fBPS(\gamma,L_{p,0};u)
\CY_{\gamma}
=\sum_{|\vv|\leq|P|}e^{(\vv,\fb)}\big( F(\fa)\big)^{|\vv|} \lim_{\xi\to0}\int_{\widetilde\CM_{KN}^\xi(P,\vv)}e^{\omega +\mu_T}\widehat{A}_T(T\widetilde\CM_{KN}^\xi)\cdot C_{T\times T_F}(\CV(\CR))~.
\end{split}\end{align}
Since the formula is valid for an infinite number of line defects, we can use it both to express $\CY_\gamma$ in terms of $\fa,\fb$ (or vice versa) and to determine relations between Dirac indices and characteristic numbers on certain Kronheimer-Nakajima spaces.

The outline of the paper will be as follows. We will begin by reviewing 't Hooft defects in 4D $\CN=2$ $G=SU(2)$ asymptotically free gauge theories with adjoint and fundamental matter . Then we will discuss the localization results for their expectation value and the connection to characteristic numbers of Kronheimer-Nakajima spaces. Then we will go on to discuss spectral networks and the relation to the index of a twisted Dirac operator on singular monopole moduli space in the semiclassical limit. Then we will discuss the comparison of localization and spectral networks  and provide an index formula for Dirac operators on singular monopole moduli space and a formula for the characteristic numbers of Kronheimer-Nakajima spaces. We will illustrate the derivation of these formulas fully in the example of  the $SU(2)$ ${\rm N}_f=0$ theory.

We expect that this technique can be extended to more general 4D $\CN=2$ theories of class $\CS$. In the more general setting, the calculations analogous to the ones presented in this paper could be used to compute more general indices of the twisted Dirac operator on singular monopole moduli space with higher rank gauge group and coupled to more complicated vector bundles with hyperholomorphic connection. 

Additionally, this paper is related to the work of \cite{Jeong:2018qpc,Nekrasov:2011bc} in which the authors investigate the generating function for certain holomorphic Darboux coordinates (what we are calling complexified Fenchel-Nielson coordinates) on the moduli space of $SL(2;\IC)$ flat connections on a punctured Riemann surface. There, the authors  relate the generating function for these coordinates to the effective twisted superpotential of the corresponding 4D $\CN=2$ theory of class $\CS$ in the presence of a $\half\Omega$-deformation and discuss its relevance to quantum Hitchin systems. 

\section{'t Hooft Defects in 4D $\CN=2$ $SU(2)$ Gauge Theories on $\IR^3\times S^1$}

\label{sec:sec2}

Our setting will be 4D $G=SU(2)$ $\CN=2$ gauge theory on $\IR^3\times S^1$ with adjoint or fundamental matter. We will be considering the expectation value of reducible 't Hooft defects which are studied in \cite{Ito:2011ea,Brennan:2018yuj,Brennan:2018rcn}.

A reducible 't Hooft defect is defined in terms of irreducible defects which are $\half$-BPS operators specified by the data $(\vx_n,P_n, \zeta)$. Here $\vx_n\in \IR^3$ specifies the insertion position, $P_n\in \Lambda_{mw}$ specifies the 't Hooft charge,\footnote{The magnetic weight lattice is defined as $\Lambda_{mw}=\{h\in \Lambda_{cochar}~|~\langle\mu_{\rm matter},h\rangle\in \IZ~,~\forall \mu_{\rm matter}\}$ where $\mu_{\rm matter}$ are highest weights specifying the representations of the hypermultiplets in the theory and the cocharacter lattice is defined as $\Lambda_{cochar}=\{h\in \ft~|~{\rm exp}\{2\pi h\}=\mathds{1}_G\}$.} and $\zeta\in U(1)$ specifies the conserved supersymmetries. Given this data, the associated 't Hooft defect is defined by imposing the local boundary conditions
\begin{align}\begin{split}
\vec{B}=\frac{P_n}{2r_n^2}\hat{r}_n+O(r_n^{-3/2})\quad, \quad X=-\frac{P_n}{2 r_n}+O(r_n^{-1/2})~,
\end{split}\end{align}
at $\vx_n$ where $\vec{r}_n=\vx-\vx_n$, $\vec{B}$ is the magnetic field, and $X$ is a real, adjoint valued Higgs field which is related to the complex Higgs field of the $\CN=2$ vectormultiplet $\Phi$ as
\be
{\rm Im}[\zeta^{-1}\Phi]=X~.
\ee
We will denote an irreducible 't Hooft defect as $L_{[P_n,0]}(\vx_n)$. We will often suppress the dependence on $\vx_n$ in addition to the phase $\zeta$.

A reducible 't Hooft defect is then defined as the product of irreducible 't Hooft defects:
\be\label{Lp0}
L_{p,0}=\left(L_{[h^1,0]}\right)^p~,
\ee
where $h^1$ is a simple magnetic weight. As an operator in a quantum theory, the expectation value of a reducible 't Hooft defect inserted at the origin is defined as the limit
\be\label{quantumred}
\langle L_{p,0}\rangle=\lim_{\vx_i\to 0}\left\langle L_{1,0}(\vx_1)... L_{1,0}(\vx_p)\right\rangle~.
\ee
Note that $L_{p,0}$ should not to confused with $L_{[ph^1,0]}$.

\subsection{Monopole Bubbling}

\label{sec:bubbling}

In order to compute the expectation value of an 't Hooft defect, it is necessary to understand the BPS field configurations. These are generically described by a collection of dyonic states in the presence of an 't Hooft defect. An important feature of the interaction between magnetically charged states  and 't Hooft defects is the phenomenon called monopole bubbling \cite{Kapustin:2006pk}.

Monopole bubbling is the process in which a smooth monopole is absorbed by an 't Hooft defect, thus screening the magnetic charge of the defect. Additionally, it gives rise to an effective SQM living on the world volume of the screened 't Hooft defect which we will refer to as the bubbling SQM. This can be understood via the following string theory construction \cite{Brennan:2018yuj,Brennan:2018moe}.

Consider embedding the 4D $\CN=2$ $SU(2)$ gauge theory on the world volume theory of a pair of parallel D3-branes which are localized at $ x^{5,6,7,8,9}=0$ and $x^4=\pm \half v$. This embedding can be achieved by adding a sufficiently large mass deformation to the adjoint hypermultiplet associated with the excitations along the $x^{6,7,8,9}$-directions.\footnote{For example by introducing an $\Omega$-deformation in the $x^{6,7,8,9}$-directions.} This seperation in the $x^4$-direction introduces a Higgs vev $X_\infty=v H_1$, where $H_1$ is a simple coroot.

\begin{figure}
\begin{center}
\includegraphics[scale=1]
{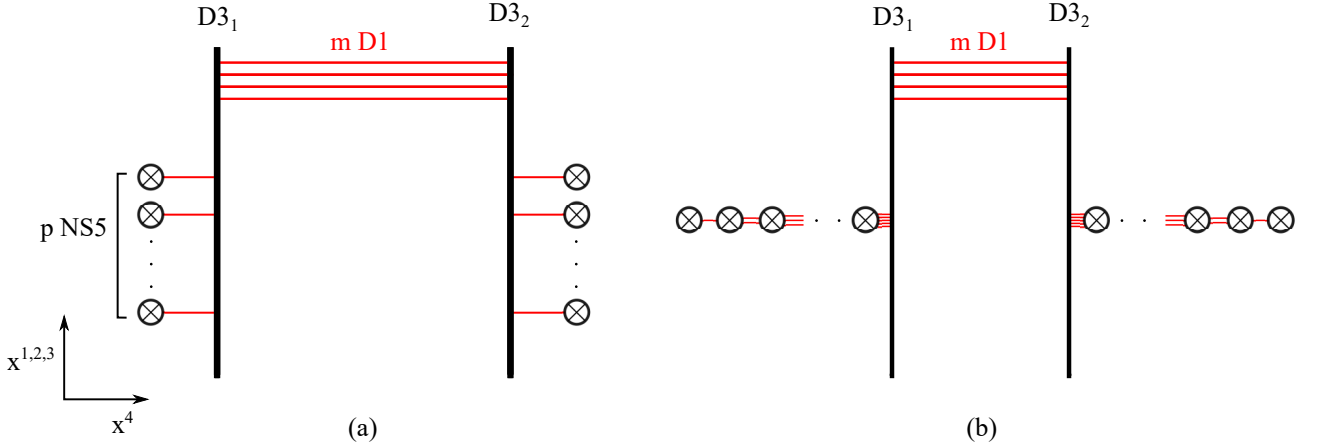}
\end{center}
\caption{In this figure we show the brane configuration describing singular monpole configurations on a stack of D3-branes (a). In (b) we show the singular limit in which we have a reducible monopole given by many spatially coincident NS5-branes in the $x^{1,2,3}$-directions. }
\label{fig:branes}
\end{figure}

Here, smooth monopoles are described by D1-branes running between the D3-branes in the $x^4$-direction \cite{Diaconescu:1996rk}. Additionally, a reducible 't Hooft defect at $\vx_n$ is described by a collection of NS5-branes that are localized at $x^{1,2,3}=\vx_n$ and spread out in the $x^4$-direction between the D3-branes. If we combine these two elements so that there are m D1-branes running between the D3-branes and $2p$ NS5-branes localized at $\vx_n$, then we have an asymptotic, relative magnetic charge and 't Hooft charge
\be
\tilde\gamma_m=\gamma_m-[P_n]^-={\rm m}\, H_1\quad, \quad P_n=2p\, \hat{h}^1~,
\ee
where $\hat{h}^1$ is a simple cocharacter (not magnetic weight), and $[P_n]^-$ is the image of the 't Hooft charge in the totally negative Weyl chamber. See Figure \ref{fig:branes}.

Here, monopole bubbling can be understood by moving D1-branes to be coincident with the collection of NS5-branes in the $x^{1,2,3}$-directions. We can then understand the bubbling SQM living on the world volume of the 't Hooft defect by performing a sequence of Hanany-Witten transformations so that the bubbled D1-branes only end on NS5-branes \cite{Brennan:2018yuj,Hanany:1996ie,Gaiotto:2008sa}.\footnote{Here Hanany-Witten transformations are represented by pulling NS5-branes through D3-branes to create a D1-brane stretched between them. See \cite{Brennan:2018yuj} for details. } In the case where $(p-k)$ D1-branes have bubbled, the theory living on the 't Hooft defect is described by a $\CN=(0,4)$ quiver SQM with quiver:

\begin{center}
\begin{tikzpicture}[
cnode/.style={circle,draw,thick,minimum size=9mm},snode/.style={rectangle,draw,thick,minimum size=9mm}]
\node[cnode] (1) {1};
\node[cnode] (2) [right=.4cm  of 1]{2};
\node[cnode] (5) [right=0.6cm of 2]{\tiny{$k-1$}};
\node[cnode] (6) [right=0.4cm of 5]{$k$};
\node[cnode] (7) [right=0.4cm of 6]{$k$};
\node[cnode] (20) [right=0.8cm of 7]{$k$};
\node[cnode] (9) [right=0.4cm of 20]{$k$};
\node[cnode] (10) [right=0.4cm of 9]{\tiny{$k-1$}};
\node[cnode] (14) [right=0.8cm of 10]{$2$};
\node[cnode] (17) [right=0.4cm of 14]{1};
\node[snode] (18) [below=0.5cm of 6]{1};
\node[snode] (19) [below=0.5cm of 9]{1};
\draw[-] (1) -- (2);
\draw[dotted] (2) -- (5);
\draw[-] (5) --(6);
\draw[-] (6) -- (7);
\draw[dotted] (7) -- (20);
\draw[-] (20) -- (9);
\draw[-] (9) -- (10);
\draw[dotted] (10) -- (14);
\draw[-] (14) -- (17);
\draw[-] (6) -- (18);
\draw[-] (9) -- (19);
\end{tikzpicture}
\end{center}

\noindent where the node $k$ is repeated $2p-2k+1$ times\footnote{In the case where $p=k$, there is a single $U(k)$ node with 2 fundamental hypermultiplets adjoined to it.} and
\be
P=2p \,\hat{h}^1\quad, \quad P-\vv=k H_1~,
\ee
where $\vv$ is the effective 't Hooft charge. We will refer to this quiver as $\Gamma(P,\vv)$.

In the case of a 4D theory with adjoint matter, the bubbling SQM is enhanced to a $\CN=(4,4)$ quiver SQM, and in the case of ${\rm N}_f$ fundamental hypermultiplets, it is modified by coupling to ${\rm N}_f$ short fundamental Fermi multiplets to the central gauge node:

\begin{center}
\begin{tikzpicture}[
cnode/.style={circle,draw,thick,minimum size=9mm},snode/.style={rectangle,draw,thick,minimum size=9mm}]
\node[cnode] (1) {1};
\node[cnode] (2) [right=.4cm  of 1]{2};
\node[cnode] (5) [right=0.6cm of 2]{\tiny{$k-1$}};
\node[cnode] (6) [right=0.4cm of 5]{$k$};
\node[cnode] (7) [right=0.4cm of 6]{$k$};
\node[cnode] (3) [right=.8cm of 7]{$k$};
\node[cnode] (20) [right=0.8cm of 3]{$k$};
\node[cnode] (9) [right=0.4cm of 20]{$k$};
\node[cnode] (10) [right=0.4cm of 9]{\tiny{$k-1$}};
\node[cnode] (14) [right=0.8cm of 10]{$2$};
\node[cnode] (17) [right=0.4cm of 14]{1};
\node[snode] (18) [below=0.5cm of 6]{1};
\node[snode] (19) [below=0.5cm of 9]{1};
\node[snode] (13) [above=0.6cm of 3]{${\rm N}_f$};
\draw[-] (1) -- (2);
\draw[dotted] (2) -- (5);
\draw[-] (5) --(6);
\draw[-] (6) -- (7);
\draw[dotted] (7)-- (3);
\draw[dotted] (3) -- (20);
\draw[-] (20) -- (9);
\draw[-] (9) -- (10);
\draw[dotted] (10) -- (14);
\draw[dashed] (13) -- (3);
\draw[-] (14) -- (17);
\draw[-] (6) -- (18);
\draw[-] (9) -- (19);
\end{tikzpicture}
\end{center}

\section{Localization}
\label{sec:localization}

We can compute the expectation value of an 't Hooft defect on $\IR^3\times S^1$ using localization. Due to the fact that we are on a non-compact space, we must introduce an IR-regulator which is accomplished via a $\half\Omega$-deformation with corresponding fugacity $\epsilon_+$. The expectation value of $L_{p,0}$ can be defined as a supersymmetric index
\be
\langle L_{p,0}\rangle={\rm Tr}_{\CH_{L_{p,0}}}(-1)^F e^{-\beta H+\epsilon_+ J_+ +i Q\cdot \Theta+m_f \cdot F}~,
\ee
where $\CH_{L_{p,0}}$ is the Hilbert space of the quantum field theory with an insertion of $L_{p,0}$, $\beta$ is the radius of the thermal circle, $H$ is the Hamiltonian, $J_+$ and $F$ are the generators of the $\half \Omega$-deformation and flavor symmetries with fugacities $\epsilon_+$ and $m_f$ respectively, $Q=(\gamma_e,\gamma_m)$ is the vector of electric and magnetic charges, and $\Theta=(\theta_e,\theta_m)$ is the vector of electric and magnetic theta angles.

Here we will be fixing the electric and magnetic theta angles $\theta_e,\theta_m$. The electric theta angle can be defined in terms of local field configurations by the holonomy of the gauge connection along the thermal circle at infinity
\be
\oint_{S^1_\infty}A_t dt=\theta_e~.
\ee
The magnetic theta angle, is similarly naturally defined as the holonomy of the dual magnetic gauge field. However, in order to fix both $\theta_e$ and $\theta_m$, we must define the magnetic theta angle as the Fourier dual of path integral with fixed magnetic charge $\langle L_{\vec{p},0}\rangle_{\gamma_m}$
\be\label{fourierTheta}
\langle L_{{p},0}\rangle_{\theta_m}=\sum_{\gamma_m} \langle L_{{p},0}\rangle_{\gamma_m} e^{-2\pi i \gamma_m \theta_m}~.
\ee
Implementing localization leads to an expectation value which is of the form of \eqref{localized}. The
details of the localization computation are contained in \cite{Ito:2011ea,Brennan:2018yuj,Gomis:2011pf}.

\subsection{Expectation Value of Line Defects}

In the case of $G=SU(2)$, as we consider here, the expectation value of an 't Hooft defect that is  computed by localization is naturally written in terms of two complexified Fenchel-Nielsen coordinates: $\fa,\fb$. These form a maximal set of holomorphic coordinates on the Hitchin moduli space $\CM_\zeta$.
 Associated to a weak coupling description, these coordinates have a semiclassical expansion
\begin{align}\begin{split}\label{absemi}
\fa=&i \theta_e-2\pi \beta Y_\infty+...~,\\
\fb=&i \theta_m+\frac{8\pi^2\beta}{g^2}X_\infty-\vartheta \beta Y_\infty+...~,
\end{split}\end{align}
where $\theta_m$ and $\theta_e$ are the magnetic and electric theta angles, $\zeta^{-1}\Phi_\infty=Y_\infty+iX_\infty$ are the real and imaginary parts of the phase rotated vev of the adjoint-valued Higgs field $\Phi$ of the $\CN=2$ vectormultiplet, and $\vartheta$ is the real part of the complex gauge coupling $\tau$. Note that we will generally take $\Phi_\infty$ to be fixed so that $\fa,\fb$ have $\zeta$-dependence via $X_\infty,Y_\infty$. 
Additionally, here $\beta$ is the radius of the thermal circle and $(...)$ correspond to non-perturbative corrections, which we will discuss later in Section \ref{sec:NPFN}. More exact expressions for them are given in Section \ref{sec:CFN}.

In these coordinates, the expectation value of  an 't Hooft defect is written \cite{Ito:2011ea,Gomis:2011pf}
\be\label{localized}
\langle L_{p,0}\rangle=\sum_{v\in \IZ_+\,:\, v\leq p}\cosh(\vv,\fb) (F(\fa))^{v}Z_{mono}(\fa,m,\epsilon;P,\vv)\quad, \quad P={\rm diag}(p,-p)~,
\ee
Here the sum is over monopole bubbling configurations labeled by the effective 't Hooft charge $\vv={\rm diag}(v,-v)$. In each summand, the contribution $\cosh(\vv,\fb) (F(\fa))^{v}$ can be attributed to the contribution of the bulk fields whereas $Z_{mono}(P,\vv)$ can be attributed to the bubbling SQM living on the world volume of the 't Hooft defect in the given bubbling configuration.

As in \cite{Brennan:2018yuj,Brennan:2018rcn}, we will focus on the contribution coming from $Z_{mono}(P,\vv)$. This contribution can be understood as the Witten index $I_W$ of the bubbling SQM as described in Section \ref{sec:bubbling}.\footnote{\label{Nf2N}There is an additional subtlety in the case of the $SU(2)$ N$_f=4$ theory (and indeed in any $SU(N)$ N$_f=2N$ theory). Here the $Z_{mono}(P,\vv)$ has the interpretation as the ground state index $I_{\CH_0}=\lim_{\beta\to \infty}I_W(\beta)$. Additionally, there are some issues that arise with applying localization to the related bubbling SQM which requires adding an additional contribution

\be
Z_{mono}:=I_{\CH_0}=I_{\CH_0}^{(Loc)}+I_{asymp}~,
\ee
where $I_{asymp}$ counts ground states on non-compact directions of field space with finite potential energy. See \cite{Brennan:2018rcn,Assel:2019iae} for more details.} These theories naturally have an action which is $\CQ$-exact:
\be
S=\CQ\cdot V~,
\ee
where $\CQ$ is a real supercharge that satisfies:
\be
\CQ^2=H+\fa Q_\fa+\epsilon_+ J_+ +m_f\cdot F~,
\ee
where $Q_\fa$, $J_+$, and $F$ are charges for the flavor symmetry corresponding to global gauge transformations, the $R$-symmetry corresponding to the $\half\Omega$-deformation, and other flavor symmetries. These generate a group of conserved symmetries of the bubbling SQMs which we will denote $T=T_G\times U(1)_{\epsilon_+}\times  T_f$ where $T_G,T_f$ are the maximal tori of the gauge and flavor group respectively.

Because the bubbling SQM is described by a $\CN=(0,4)$ quiver SQM, the path integral naturally localizes to a $T$-equivariant integral over the vacuum moduli space.
 In order to use localization, we must introduce an FI-parameter in the SQM. This lifts the Coulomb and mixed branches and causes the path integral to localize to a finite dimensional ($T$-equivariant) integral over the Higgs branch which is isomorphic to the Kronheimer-Nakajima space/quiver moduli space corresponding to the quiver defining the bubbling SQM: $\Gamma(P,\vv)$.

It will be more convenient in the following discussion however to identify the quiver $\Gamma(P,\vv)$ by its gauge and hypermultiplet nodes. That is to say we will define $\Gamma(\vk,\vw):=\Gamma(P,\vv)$ where $\vk=(k_1,...,k_n)$ encode the gauge nodes and $\vw=(w_1,...,w_n)$ encode the fundamental hypermultiplet nodes.
In this notation, we can identify the Higgs branch of the bubbling SQMs with the Kronheimer-Nakajima space $\CM_{KN}(\vk,\vw)$ associated to the quiver $\Gamma(\vk,\vw)$.  $\CM_{KN}(\vk,\vw)$ is defined as the hyperk\"ahler quotient with respect to the moment maps
\begin{align}\begin{split}\label{eq:moment}
&\mu_\IC=\sum_i\sum_{j=i-1,i}B_{j,j+1} B_{j+1,j}+I_iJ_i~,\\
&\mu_\IR=\sum_{i}[B_{i,i+1}^\dagger, B_{i,i+1}]-[B_{i-1,i}^\dagger,B_{i-1,i}]+I_i   I^\dagger_i-J^\dagger_i J_i~,
\end{split}\end{align}
where $V_i\cong \IC^{k_i}$ and $W_i\cong \IC^{w_i}$ with
\be
B_{j,j+1}:V_{j}\to V_{j+1}\quad, \quad B_{j+1,j}:V_{j+1}\to V_{j}\quad, \quad I_i:V_i\to W_i\quad, \quad J_i:W_i\to V_i~,
\ee
and the gauge group $G=\prod_i U(k_i)$  acts as
\be
B_{i,i+1}\mapsto g_i^{-1}B_{i,i+1}g_{i+1}\quad, \quad I_i\mapsto Ig_i\quad, \quad J_i\mapsto g_i^{-1}J_i\quad, \quad g_i\in U(k_i)~.
\ee
Physically, the $B_{i,i\pm1}$ can be interpreted as the bosonic fields of the bifundamental hypermultiplets and the $I_i,J_i$ can be interpreted as the bosonic fields of the fundamental hypermultiplets.

\subsection{$Z_{mono}$ in terms of Characteristic Numbers}

\label{sec:charnum}

Thus far we have not taken into account the varying matter content of the 4D theories. Coupling the 4D theory to matter hypermultiplets gives rise to a vector bundle, which we will call the \emph{matter bundle}, over the same moduli space $\CM_{KN}(\vk,\vw)$. This modifies the $T$-equivariant integral so that we can identify $Z_{mono}(P,\vv)$ with the $T$-equivariant integral over $\CM(\vk,\vw)$ of some characteristic class

%
\be
Z_{mono}(P,\vv)\sim \int_{\CM_{KN}(\vk,\vw)} e^{\omega +\mu_T}~char_T(T \CM_{KN})~,
\ee
where $\omega$ is the real symplectic (1,1)-form in some choice of complex structure (which we will define shortly), $\mu_T$ is the moment map for the action of $T$, and   $char_T(T \CM_{KN})$ is a $T$-equivariant characteristic class on $T\CM_{KN}$ depending on the matter content of the theory.

This equation is not strictly well defined since $\CM_{KN}(\vk,\vw)$ is a singular space. However, the singularities can be resolved by introducing an FI-parameter $\xi$ \cite{Kronheimer:1990}.  We will denote the resolved space $\widetilde\CM_{KN}^\xi(\vk,\vw)$.

In 4D, the parameter $\xi$ comes with a choice of direction of on $\IR^3$ (and consequently on the resolved space $\widetilde\CM_{KN}^{\xi})$ which corresponds to deforming the product 
\be\label{eq:res}
\Big\langle L_{p,0}(0)\Big\rangle\longrightarrow\Big\langle L_{1,0}(\vx_1) ~L_{1,0}(\vx_2)~...~L_{1,0}(\vx_p)\Big\rangle\quad,\qquad \vx_i-\vx_{i+1}=\xi \hat{z}~,
\ee
in that direction (we have chosen the $\hat{z}$-direction).
The choice of direction defines a complex structure on $\widetilde\CM_{KN}^{\xi}(\vk,\vw)$ and it is in this complex structure that $\omega$ is the (1,1) symplectic form.
Note that by comparing to the definition of the reducible 't Hooft defect \eqref{quantumred}, computing the expectation value $\langle L_{p,0}\rangle$ requires taking the limit $\xi\to 0$.
See \cite{Brennan:2018yuj} for more details. 

%

Now we can use the result of \cite{Brennan:2018yuj,Ito:2011ea} for the characteristic class for the 4D asymptotically free $SU(2)$ theories with fundamental and adjoint matter to write the monopole bubbling contribution
\begin{align}\begin{split}\label{GeneralChar}
Z_{mono}(P,\vv)&=
\lim_{\xi \to 0`} \int_{\widetilde\CM_{KN}^{\xi}(\vk,\vw)} e^{\omega+\mu_T}\begin{cases} \widehat\chi_y^T(T\widetilde\CM_{KN})& \CN=2^\ast\text{ theory}\\
\widehat{A}^T(T\widetilde{\CM}_{KN})\cdot C_{T\times T_F}(\CV_{{\rm N}_f})& {\rm N}_f \text{ theory}\end{cases}
\end{split}\end{align}
where $\widehat\chi_y^T$ the $T$-equivariant $\widehat\chi_y$-genus with $y=e^{-m+\epsilon_+}$, and $\CV_{{\rm N}_f}\to \widetilde\CM_{KN}$ is a vector bundle with
\be
ch(\CV_{{\rm N}_f})=\sum_i e^{x_i}\quad,\qquad C_{T\times T_F}(\CV_{{\rm N}_f})=\prod_i \left(e^{x_i}-e^{-x_i}\right)~.
 \ee
 See \cite{Moore:1997dj,Moore:1998et,Nekrasov:2002qd,Ito:2011ea,Brennan:2018moe,Martens:2006hu,PratoWu} for details on computing these integrals. 


\section{Spectral Networks}
\label{sec:3}

In this section we will review the technology of spectral networks
 \cite{Gaiotto:2012rg,Gaiotto:2012db}.  
 Spectral networks are a method for computing the the holonomy of flat connections on a Riemann surface, which, as discussed previously, can be identified with the expectation value of a line defect $L$ in the class $\CS$ construction.
It provides a set of ``Darboux coordinates'' $\CY_\gamma$
%
on $\CM_{flat}(C;G_\IC)$, the moduli space of flat $G_\IC$ connections on $C$ associated to $\gamma\in H_1(\Sigma;\IZ)$.
Spectral networks naturally computes the expectation value of the 't Hooft defect as a Laurent series in these coordinates
\be\label{eq:Dexp}
\langle L\rangle=\sum_{\gamma\in H_1(\Sigma;\IZ)} \fro(\gamma;L_\CP)\CY_\gamma~,
\ee
whose coefficients are framed BPS indices.

\subsection{Theories of Class $\mathcal{S}$ }

\label{sec:classS}

Theories of class $\CS$ are those which are constructed by taking the six-dimesional $\CN=(2,0)$ theory and compactifying it along an oriented Riemann surface $C$ with a topological twist \cite{Klemm:1996bj,Gaiotto:2009hg,Gaiotto:2009we,Witten:1997sc}. For the type $SU(N)$ theories of class $\CS$, this can be described as the low energy effective theory of a stack of $N$ M5-branes wrapped on $C\times M_4$ with the same topological twist where $M_4$ is 4D spacetime. When $M_4 = \mathbb{R} \times S^1$ where $S^1$ has radius $R$ 
the space of quantum vacua can be identified with the moduli space of solutions to the  Hitchin equations on $C$
\be
F_C+R^2[\varphi,\bar \varphi]=0\quad,\qquad \bar\partial_{A_C}\varphi=0~,
\ee
with gauge group $G=SU(N)$ where $A_C$ is the $G$-connection. 
 Given a solution of these equations, we can identify the Seiberg-Witten curve and differential as
\be
\Sigma=\{\text{ det}(xdz-\varphi)=0\}\subset T^\ast C\quad,\qquad \lambda_{SW}=x dz~,
\ee
where $(x,z)\mapsto xdz$ are coordinates on $T^\ast C$. Intuitively, the Coulomb branch 
vacuum is described by separating multiple M5-branes wrapped on $C$  along a transverse direction
to produce a single $M5$ brane wrapped on Riemann surface $\Sigma$ which is an  $N$-branched cover $\Sigma\to C$.

\subsubsection{Labeling Line Defects In Class $\CS$}
\label{sec:CFN}
 
As described in Section 7 of \cite{Gaiotto:2010be} a natural class of line defects in class $\CS$ theories is obtained by 
considering a semi-infinite M2-brane that ends on a one-dimensional submanifold $\CP \subset C$ times a line in $M_4$. 
The line defect is labeled by a representation $\CR_\alpha$ of $SU(N)$ associated to each connected component $\CP_\alpha$ 
of $\CP$ and a phase $\zeta$, where $\zeta$ determines the unbroken supersymmetry. (Geometrically $\zeta$ describes how the 
M2-brane extends in the extra dimensions.) We can denote the resulting line defect by $L(\CR,\CP, \zeta)$.
We now consider the case where the line defect wraps the circle of radius $R$ in $M_4$ at a fixed point in $\mathbb{R}^3$. 
In this case the vacuum is described by a solution to Hitchin's equation to which  
we can associate a  flat $G_\IC$ connection
\be
\CA=R\zeta^{-1} \varphi+A_C+R\zeta \bar \varphi ~ . 
\ee
Because these theories are partially topologically twisted, the expectation value of such a 
line defect can be expressed in terms of the holonomy of this flat gauge field  \cite{Gaiotto:2009hg,Gaiotto:2010be}:
\be\label{4point5}
\langle L_\CP\rangle=\prod_{\alpha} {\rm Tr}_{\CR_\alpha}{\rm ~Hol}_{\CP_\alpha}  \,\CA~.
\ee

In this paper we will be focusing on the case of theories of class $\CS$ with gauge group $SU(2)$. 
We will take all the representations $\CR_{\alpha}$ to be the fundamental representation. Therefore, 
the line defects can be labeled by $L(\CP, \zeta)$ where $\CP$ is a smooth one-dimensional 
submanifold of $C$. Isotopy classes of such submanifolds can be conveniently labeled, given a 
pants decomposition of $C$ in terms of Dehn-Thurston parameters: \footnote{The importance of being careful about connected components 
in the Dehn-Thurston theorem was first made clear to us in joint work 
with Anindya Dey while checking predictions of S-duality in class S theories 
with gauge group $G=SU(2)$.} \\

\textbf{Theorem (Dehn-Thurston): \cite{Dehn,Thurston}} Let $C$ be an oriented Riemann surface with negative Euler characteristic that has genus $g$ and $n$ punctures. Let $\{\gamma_i\}_{i=1}^{3g-3+n}$ be a maximal set of non-intersecting curves defining a pants decomposition of $C$ and let $\{\gamma_i\}_{i=3g-3+n+1}^{3g-3+2n}$ be a collection of simple closed curves near the punctures. There is a mapping
\begin{align}\begin{split}
D:\CI(C)&\to \IZ_{\geq0}^{3g-3+2n}\times \IZ^{3g-3+2n}~, \\
\gamma&\mapsto \big(\langle \gamma, \gamma_i\rangle,\vec{q}\big)
\end{split}\end{align}
where $\CI(C)$ is the set of isotopy classes of closed one dimensional submanifolds, $q_i$ is the twisting number with respect to $\gamma_i$, and $\langle~,~\rangle$ is the intersection number. Elements in the image of $D$ are denoted $(\vec{p},\vec{q})$ and are called \textit{Dehn-Thurston parameters.}\\

The choice of  $\{\gamma_i\}_{i=1}^{3g-3+n}$ above correspond to a weak coupling decomposition of the UV curve $C$, and 
specifies a Lagrangian duality frame with gauge algebra $\mathfrak{s}\mathfrak{u}(2)^{\oplus h}$ with $h=3g-3+n$. 
 Each curve corresponds to a weakly coupled $SU(2)$ gauge group in the 4D theory.

Now consider the line defect associated to a generic 1D submanifold $\gamma_{\vec{p},\vec{q}}$ with Dehn-Thurston (DT)  parameters $(\vec{p},\vec{q})=(p_1,...,p_h,q_1,...,q_h)$. This submanifold will have a set of connected components $\gamma_{\vec{p},\vec{q}}=\bigoplus_{\alpha=1}^k \gamma^{(\alpha)}_{\vec{p},\vec{q}}$ labeled by $\alpha$, 
each of which has its own Dehn-Thurston parameters: $(\vec{p}^{(\alpha)},\vec{q}^{(\alpha)})=(p_1^{(\alpha)},...,p_h^{(\alpha)},q_1^{(\alpha)},...,q_h^{(\alpha)})$. The line defect $L(\gamma_{\vec{p},\vec{q}}, \zeta)$ then decomposes as a product of line defects
\be
L_{\gamma_{\vec{p},\vec{q}}}=\prod_{\alpha=1}^k L_{\gamma_{\vec{p},\vec{q}}^{(\alpha)}}~,
\ee

In \cite{Drukker:2009tz} it is conjectured that the line defects $L(\CP,\zeta)$ are 
the same as the 't Hooft-Wilson line defects of the Lagrangian theory with gauge 
algebra $\mathfrak{s}\mathfrak{u}(2)^{\oplus h}$. Moreover, it is proposed that 
the Dehn-Thurston parameters should be identified with the 't Hooft-Wilson parameters
characterizing the magnetic and electric charges. This cannot be true in general, 
but it seems highly plausible for those Dehn-Thurston parameters that correspond 
to one-dimensional submanifolds $\gamma_{\vec p, \vec q}$ with only one connected 
component. In this case the proposal of Drukker-Morrison-Okuda is that $L(\CP,\zeta)$ 
corresponds to the 4D line operator $L_{[P^{(i)},Q^{(i)}]}$ which has  't Hooft-Wilson charges
\be\label{PQdecomp}
P=\bigoplus_{j=1}^h p_j  h^{I(j)}
\quad,\quad Q=\bigoplus_{j=1}^h q_j \lambda^{I(j)}~,
\ee
where $h^{I(j)}$ is the simple magnetic weight, $\lambda^{I(j)}$ is the simple weight of the $j^{th}$ factor of the gauge group, and $h=3g-3+n$.
It should be stressed that some more work is needed to make use of this conjecture: In mathematics \emph{it is not known 
what conditions one should put on the Dehn-Thurston parameters $(\vp, \vq)$ in order for $\gamma_{\vp, \vq}$ to have a single connected component!}
The only case where this is known is the once-punctured torus (corresponding to the $G=SU(2) $ $\CN=2^*$ theory) and the four-punctured sphere (corresponding to the $G=SU(2)$ N$_f=4$ theory) \cite{FLuo}. 
In that case there are only a pair of DT parameters $(p,q)$ and $\gamma_{(p,q)}$ has $g$ connected components, where $g$ is the gcd of $p$ and $q$.

%
%

More generally, in the case where the four-dimensional gauge group is $G=SU(2)$ we have only a pair of DT parameters $(p,q)$. Here the minimally charged 't Hooft defect corresponds to the line with DT parameters $(1,0)$
\be
L_{\gamma_{(1,0)}}=L_{[h^1,0]}~,
\ee
which can be identified with the highest weight representation $R_{h^1}$ of $SU(2)^\vee$.
Following the decomposition above, a line defect corresponding to DT parameters $(p,0)$ is the $p^{th}$ power of the simple 't Hooft defect
\be
L_{\gamma_{(p,0)}}=\left(L_{[h^1,0]}\right)^p~.
\ee
Thus, we see that the `t Hooft defect corresponding to $L_{\gamma_{(p,0)}}$ is reducible. 
By equation \eqref{4point5} the vev is the trace of the holonomy in the representation $R_{h_1}^{\otimes p}$. 
This is the origin of our notation from \eqref{Lp0}
\be
L_{p,0} := L_{\gamma_{(p,0)}}~.
\ee
  By contrast $L_{[ph^1,0]}$ 
corresponds to a trace in the representation $R_{ph^1}$.  If the vev of $L_{p,0}$ is expressed as 
$(2 \cos \theta)^p$ then the vev of $L_{[ph^1,0]}$ has vev $\frac{\sin\big((p+1)\theta\big)}{\sin\theta}=U_p(\cos\theta)$ 
where $U_p$ is a Tchebyshev polynomial. 


\subsubsection{Complexified Fenchel-Nielsen Coordinates}
\label{sec:CFN}

Because the expectation values of line defects in theories of class $\CS$ are given by the trace of the holonomy of a flat connection,
 they are holomorphic functions on  Seiberg-Witten moduli space.
This allows them to be expressed via the AGT correspondence in terms of a particular set of holomorphic coordinates called complexified Fenchel-Nielsen coordinates. These can be defined as follows.

Choose a weak coupling region of the Coulomb branch. This defines a complex structure and comes with a maximal set of non-intersecting curves $\{\gamma_i\}_{i=1}^{3g-3+n}$ that are not isotopic to punctures on the UV curve $C$ which correspond to weakly coupled gauge groups indexed by $i$. \footnote{Here we are restricting to the case of Lagrangian theories of class $\CS$ with $SU(2)$ gauge group. } Associated to each $\gamma_i$, we can define the holomorphic  coordinates $\{\fa_i\}\in \ft_\IC$ defined by
\be
\langle L_{\gamma_i}\rangle={\rm Tr}_N e^{\fa_i}~.
\ee
The $\{\fa_i\}$ are Poisson commuting with respect to the standard, symplectic $(2,0)$-form $\Omega_J$ on Seiberg-Witten moduli space \cite{Dimofte:2011jd,Kapustin:2006pk}
\be
\Omega_J\left(\frac{\partial}{\partial \fa_i},\frac{\partial}{\partial \fa_j}\right)=0~,
\ee
and form a maximal set of Poisson commuting holomorphic functions. 

Now we can define a set of symplectically dual coordinates $\{\fb_i\}\in \ft_\IC$ with respect to $\Omega_J$ such that
\be\label{symplecticFN}
\Omega_J=\frac{1}{\hbar}\sum_i {\rm Tr}_N(d\fa_i\wedge d\fb_i)~.
\ee
We can then fix the redundancy $\fb_i \to \fb_i+f_i(\fa)$ where $\partial_{\fa_i}f_j=\partial_{\fa_j}f_i$ by specifying the semiclassical limit as in \eqref{absemi}.

In the case of a single $SU(2)$ gauge group, as we consider here, the above discussion reduces to a single pair of Fenchel-Nielsen coordinates $\fa,\fb$. These define the Fenchel-Nielsen coordinates $\fa,\fb$ used for localization in Section \ref{sec:localization}. 

\subsection{Spectral Networks in Theories of Class $\CS$}

For this paper, we are only considering the case of spectral networks in theories of class $\CS$ of type $G=SU(2)$, $G_{\IC}=SL(2;\IC)$. 
However, the following discussion generalizes to all ADE-type gauge groups \cite{Longhi:2016rjt,Longhi:2016bte,Gaiotto:2010be,Gaiotto:2012rg,Gaiotto:2012db}.

A \textit{spectral network} $\CW$ subordinate to the covering $\Sigma\to C$ 
is a collection of oriented, open paths $w$ on $C$ called \textit{walls} that begin at branch points and flow to punctures of $C$ or other branch points.
We will be interested in a  special class of spectral networks which arise naturally in theories of class $\CS$  called \textit{WKB spectral networks}. These are defined by a meromorphic, quadratic differential $\varphi_2$ on the closure $\overline{C}$ of $C$ and a choice of $\vartheta\in \IR/2\pi \IZ$.  \footnote{For our case we will want to pick $e^{i \vartheta}=\zeta$ where $\zeta$ is the phase of the line defect.} Locally, the quadratic differential can be written
\be
\varphi_2=u(z) (dz)^2 ~.
\ee
This can be used to define a foliation of $C$ by curves $\gamma$ which satisfy
\be
e^{-2i \vartheta}u(\gamma(t))\left(\frac{d\gamma}{dt}\right)^2\in \IR_+~,
\ee
where where $t$ is an affine parameter for $\gamma$. The corresponding  spectral network $\CW(\varphi_2,\vartheta)$ is then defined by the critical graph of the foliation defined by $\varphi_2$ and $\vartheta$ --- i.e. the set of limiting curves that divide the foliation into distinct sectors.

The spectral network technology developed by \cite{Gaiotto:2012rg,Gaiotto:2012db} provides a trivialization of the $SL(2;\IC)$ vector bundle over the complement $C\backslash \CW$ and gives gluing conditions across the walls. 
This trivialization is subordinate to the covering $\Sigma\to C$ such that we can locally equate the space of flat $SL(2;\IC)$ connections on $C$ to the space of flat $GL(1,\IC)$ connections on $\Sigma$
\be\label{eq:flatequiv}
\CM_{flat}(\Sigma,GL(1;\IC))\cong \CM_{flat}(C,SL(2;\IC))~,
\ee
which allows us to compute holonomies of the non-abelian vector bundle $E\to C$ in terms of holonomies of the connection of a flat line bundle on $\CL\to \Sigma$.

The moduli space $\CM_{flat}(\Sigma,GL(1;\IC))$ has  a natural set of coordinates:
\be
\CY_\gamma=\text{Hol}_\gamma \nabla^{ab}\in \widehat\IC^\ast \quad,\qquad \forall [\gamma]\in H_1(\Sigma;\IZ)~,
\ee
where $\nabla^{ab}$ is the connection on $\CL$.
These coordinates follow the multiplication rule
\be
\CY_\gamma\CY_{\gamma^\prime}=(-1)^{\langle\gamma,\gamma'\rangle}\CY_{\gamma+\gamma^\prime}~,
\ee
where $\langle~,~\rangle$ is the oriented intersection pairing. Further, the $\CY_\gamma$ satisfy
\be\label{eq:branch}
\CY_{\gamma_b}=-1\quad,\qquad \CY_{\gamma+\omega^\ast \gamma}=1~,
\ee
where $\gamma_b$ is a small loop around a branch point $b$ and $\omega:\Sigma\to \Sigma$ is the map that exchanges sheets of the covering $\Sigma\to C$. Let us define $\Sigma^\prime=\Sigma\backslash \{\text{ramification points}\}$. For generic $\CW$, we can fix a basis of $\{\gamma_i\}\in H_1(\Sigma^\prime;\IZ)\slash\langle \gamma+\omega^\ast \gamma\rangle$ to form our coordinate system on $\CM(\Sigma,GL(1))$. These $\CY_{\gamma_i}$ are the Darboux coordinates related to the spectral network $\CW$.

The holonomy of the flat non-abelian gauge connection along a 1-cycle $\CP$ can now be given in terms of $\CY_\gamma$ associated to the decomposition of $\CP$
 with respect to a basis of open paths on $C\backslash \CW$. These open paths must then be joined across the walls of the spectral network by transition functions such that the
 complete expression for the holonomy is given by the product of the holonomies along these open paths connected by transition matrices between the different regions of $C\backslash \CW$.

The holonomy of an open path in a single connected region of $C\backslash \CW$ can be written as
\be
D_{\CP}=\left(\begin{array}{cc}
\CY_\CP&0\\0&\CY^{-1}_{\CP}
\end{array}\right)\quad~\text{ or}\qquad \tilde{D}_{\CP}=\left(\begin{array}{cc}0&\CY_{\CP}\\-\CY^{-1}_{\CP}&0\end{array}\right)~,
\ee
where $D_\CP$ corresponds to the path across a simple open region and $\tilde{D}_\CP$ corresponds to a path that crosses a branch cut. Additionally, by making a convenient choice of trivialization, we can write the holonomy across a generic wall \cite{Gaiotto:2010be,Gaiotto:2012rg,Gaiotto:2012db}
\be
\CS_w=\begin{cases}
\left(\begin{array}{cc}1&S_w\\0&1\end{array}\right)&\text{ for $w$ of type 21,}\\
\left(\begin{array}{cc}1&0\\S_w&1\end{array}\right)&\text{ for $w$ of type 12,}
\end{cases}~,
\ee
where $S_w$ is some function of $\CY_{\gamma_i}$.

The case where a path crosses a double wall is a bit more subtle.
To compute the parallel transport across a double wall, one must infinitesimally displace the phase $\zeta$ (requires picking a resolution convention) so that the double wall is replaced by a pair of generic walls; then one can compute the holonomy across the ``double wall'' by using the rules above. This prescription allows one to compute the holonomy of a complexified flat gauge connection along any path in terms of Darboux coordinates defined by the spectral network (spectral coordinates).

In addition to these rules, there are also several consistency conditions that restrict the number of free spectral coordinates. These come from abelian gauge symmetry on open path segments and from imposing monodromy conditions around branch points and punctures. This gauge symmetry acts by rescaling the spectral coordinate $\CY_{\gamma_{ij}}$ by a function corresponding to the end points of the curve $\gamma_{ij}$
\be
\CY_{\gamma_{ij}}\to  g_i\CY_{\gamma_{ij}}g^{-1}_j~,
\ee
where the beginning and end points of $\gamma_{ij}$ are labeled by $i$ and $j$ respectively. Note that the trace of the holonomy around closed paths are invariant under such gauge transformations.

The consistency conditions we impose for monodromy around a branch point $b$ and puncture $p$ is that
\be
\text{Hol}_{\gamma_b}\nabla=\left(\begin{array}{cc}
-1&0\\0&-1
\end{array}\right)\quad,\qquad \text{Hol}_{\gamma_p}\nabla=\left(\begin{array}{cc}\CY_{\gamma_p}&0\\0&\CY_{\gamma_p}^{-1}\end{array}\right)~,
\ee
which come from the condition in \eqref{eq:branch} and the trivialization of the vector bundle at the punctures given in the data specifying the spectral network. Here $\nabla$ is the connection on $E\to C$ which locally can be related to the connection on $\CL$ as explained in \cite{Gaiotto:2012rg,Gaiotto:2012db}.

In the class $\CS$ construction, there is a natural choice of quadratic differential, given by the square of the Seiberg-Witten differential 
\be
\varphi_2=\lambda_{SW}^2~.
\ee

%
%

In this setting, we have that the charge lattice of the 4D theory is given by  $\Gamma\cong H_1(\Sigma';\IZ)$. Here there is a natural central charge function
\be
Z:\Gamma\to \IC~,
\ee
given by
\be
Z:\gamma \mapsto \int_\gamma \lambda_{SW}~.
\ee
This allows us to express the corresponding Darboux coordinate as a function of $\gamma$ as \cite{Gaiotto:2008cd}
\be \label{eq:NPDarboux}
\log\, \CY_\gamma=\frac{\pi R}{\zeta}Z_{\gamma}+\pi R\, \zeta\,\overline{Z}_\gamma+i \Theta \cdot Q_\gamma+\left\{\begin{array}{c}\text{non-perturbative}\\\text{in }g\end{array}\right\}~,
\ee
where 
 $\Theta\cdot Q_\gamma$ is the Cartesian product of the vector of electric and magnetic theta angles with the vector of electromagnetic charges associated to $\gamma$. It is important to note that these coordinates generically have non-perturbative corrections to the semiclassical expression which, while complicated, are known and given in explicit formulas in \cite{Gaiotto:2008cd}.\footnote{The leading terms were first worked out in unpublished work by B. Pioline and A. Neitzke and in unpublished work by F. Denef and G. Moore. The main point of \cite{Gaiotto:2008cd} was to give the full non-perturbative answer.}
\subsection{Wall Crossing in Spectral Networks}

An important feature of spectral networks is that they give us an excellent tool for understanding wall crossing. In this setting, wall crossing is realized by changes of topology of the spectral network $\CW(\varphi_2,\vartheta)$ as we scan the phase $\zeta=e^{i \vartheta}$ which can be lifted to $\hat\zeta\in\IC^\ast$.


The locations of the critical phases $\zeta=\zeta_c$ where the spectral network undergoes topology changes lift to a co-dimension-1 ``walls'' in $\IC^\ast$ called and are called $\CK$-walls \cite{Gaiotto:2012rg}.  Physically, each $\CK$-wall corresponds to a co-dimension loci  where $\zeta$ is aligned or anti-aligned with the phase of $\overline{Z_{\gamma_k}}$.
Here, the change in topology of the spectral network causes the Darboux coordinates to undergo a cluster-like transformation/mutation \cite{Gaiotto:2010be}
 \be
K_{\gamma_k}\,:\, \CY_{\gamma_i}\mapsto
(1-\sigma(\gamma)\CY_{\gamma_k})^{-\langle \gamma_k,\gamma_i\rangle \Omega(\gamma_k)} \CY_{\gamma_i}\quad, \quad \gamma_k\in \Gamma~,
\ee
where 
\be
\sigma(\gamma)=(-1)^{\langle \gamma_e,\gamma_m\rangle}~,
\ee
is a particular choice of quadratic refinement with respect to a choice of splitting of the charge lattice and $\gamma=\gamma_e\oplus\gamma_m$ \cite{Gaiotto:2010be}.\footnote{We will be working in the semiclassical limit so that there is always an almost canonical choice of charge lattice splitting.}

\begin{figure}[t]
\centering
\includegraphics[scale=1]
{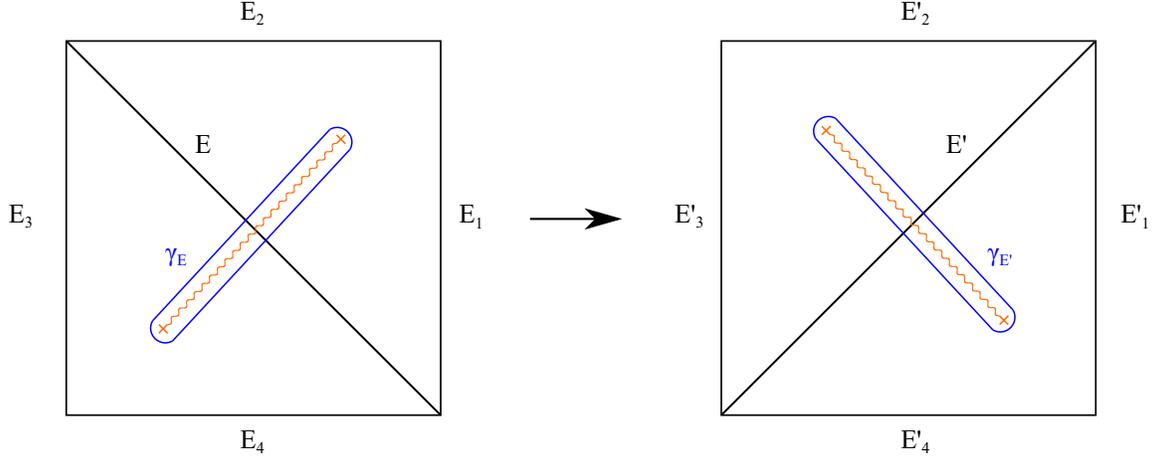}
\caption{This figure shows the flip of an edge in a triangulation (left flips to right) giving rise to a Fock Goncharov (shear) coordinate inside a quadrilateral with edges $E_1,E_2,E_3,E_4$. This figure also demonstrates the projection of the paths in $\Sigma\to C$ corresponding to the Darboux coordinates $\CY_E$ and $\CY_{E'}$. }
\label{fig:flip}
\end{figure}

However, since the expectation value of a line defect $L_\CP$ is defined by a path $\CP\subset C$ which is independent of the topology of the spectral network, the expectation value
\be
\langle L_\CP\rangle=\sum_{\gamma \in \Gamma}\fro(\gamma,L_\CP)\CY_\gamma~,
\ee
is wall crossing invariant. This means that the $\CY_\gamma$ undergo coordinate transformations which exactly cancel the wall crossing of the framed BPS indices. Thus, by studying the wall crossing properties of the $\CY_\gamma$, one can infer the wall crossing of framed BPS states.


A nice feature of generic WKB spectral networks is that the walls provide an ideal triangulation of $C$. In these networks, the associated Darboux coordinates have a natural identification with the edges of the triangulation. These coordinates are given by the holonomy along the lift under the projection $\pi:\Sigma\to C$ of a path running between the branch points of different triangles through a given edge of the triangulation. See Figure \ref{fig:flip}.   We will use the notation where the Darboux coordinate associated to the edge $E$ is denoted  $\CY_E$.


In such spectral networks, the fundamental topology shift that occurs in wall crossing is a flip of the triangulation . See Figure \ref{fig:flip}. Explicitly, in a generic WKB spectral network and consider a quadrilateral with edges $E_1,E_2,E_3,E_4$ with diagonal edge $E$, a  flip on the edge $E\mapsto E'$ acts on the corresponding Darboux coordinates by:
\begin{align}\begin{split}\label{eq:flip}
\CY_E&\mapsto \CY_{E'}^{-1}\qquad\qquad\qquad\qquad~,\quad
\CY_{E_1}\mapsto \CY_{E_1'}=\CY_{E_1}(1+\CY_E)~,\\
\CY_{E_2}&\mapsto \CY_{E_2'}=\CY_{E_2}(1+\CY_E^{-1})^{-1}\quad, \quad
\CY_{E_3}\mapsto \CY_{E_3'}=\CY_{E_3}(1+\CY_E)~,\\
\CY_{E_4}&\mapsto \CY_{E_4'}=\CY_{E_4}(1+\CY_E^{-1})^{-1}\quad,
\end{split}\end{align}
where the signed intersection pairing of the edges is $\langle E,E_i\rangle=(-1)^i$.
\begin{figure}[t]
\centering
\includegraphics[scale=0.9]
{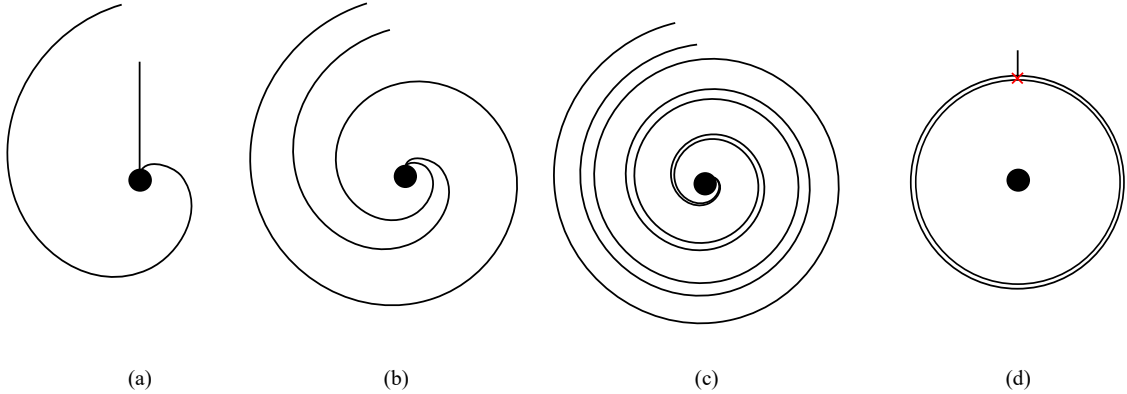}
\caption{This figure shows how a wall running to a puncture (a) twists around the puncture in a sequence of flips (b,c) and approaches the juggle in which the wall runs completely around the puncture (d). }
\label{fig:jugglepic}
\end{figure}

In the case of theories with vectormultiplets, spectral networks can also undergo a topology change called a juggle.\footnote{There is another transformation called a ``pop'' which has to do with changing the decoration of a given puncture, but this will not be important for our story. See \cite{Gaiotto:2009hg} for more details.} This can be understood as an infinite sequence of flips involving a puncture that has the effect of twisting a wall that runs to a puncture until it completely encircles it \cite{Gaiotto:2009hg}. See Figure \ref{fig:jugglepic}.

\begin{figure}[t]
\centering
\includegraphics[scale=1]
{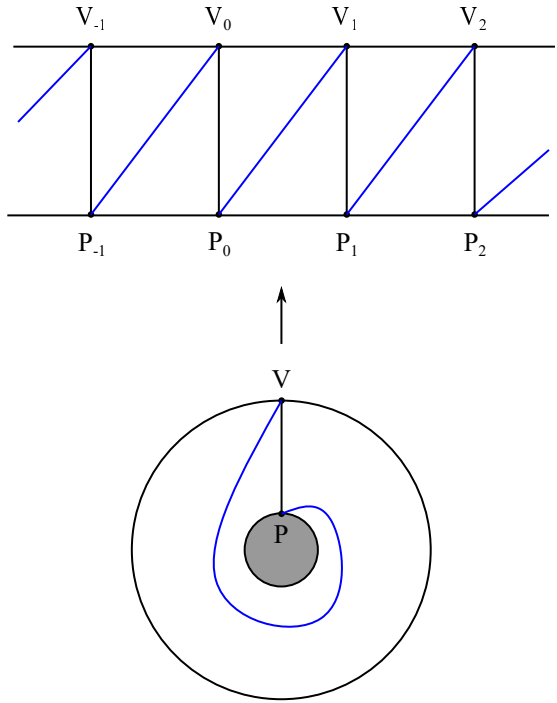}
\caption{This figure shows how to lift a spectral network on an annulus surrounding a puncture on $C$ to the simply connected cover. The puncture $P$, and exterior vertex $V$ lift an an infinite number of seperated points, denoted $\{P_i\}$ and $\{V_i\}$ respectively, connected by interior edges giving a triangulation of the strip. }
\label{fig:liftedAnn}
\end{figure}

The juggle can be understood as follows \cite{Gaiotto:2009hg}.
Consider an annulus surrounding a puncture, $P$ (which we replace by a disk with a marked point), with a single vertex $V$ of the triangulation on the outer boundary. Now consider lifting the configuration to the simply connected cover which is a triangulated infinite strip as in Figure \ref{fig:liftedAnn}. In this covering there are an infinite number of images of the interior marked point ($P\to \{P_i\})$, exterior vertex $(V\to \{V_i\})$, and edges indexed by $i\in \IZ$. We can define Darboux coordinates on the annulus as the Darboux coordinates on the triangulated strip corresponding to the different edges in the same preimage under the projection to the annulus.

\begin{figure}[t]
\centering
\includegraphics[scale=0.85]
{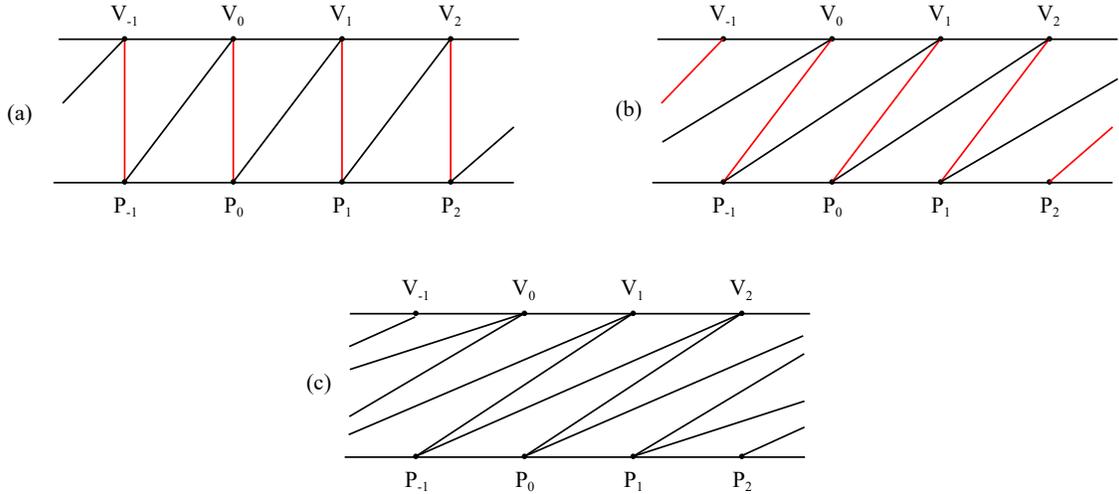}
\caption{This figure demonstrates how flips in the spectral network on the annulus corresponds to increasing winding number by considering the flips of all of the preimages in the triangulated strip. Here the processes of going from $(a)\to (b)$ and $(b)\to (c)$ requires a sequence of 2 flips where the red edges undergo the flip. }
\label{fig:juggle}
\end{figure}

 If we choose an ordering of the lifted images of the vertices, we can define a winding number of an interior edge by the difference of the image number of the end points. Further, we can iteratively increase (decrease) the winding numbers of the interior edges by performing a sequence of simultaneous flips on all of the preimages of the  the interior edge with the lowest (highest) winding number. See Figure \ref{fig:juggle}.

\begin{figure}[t]
\begin{center}
\includegraphics[scale=1]
{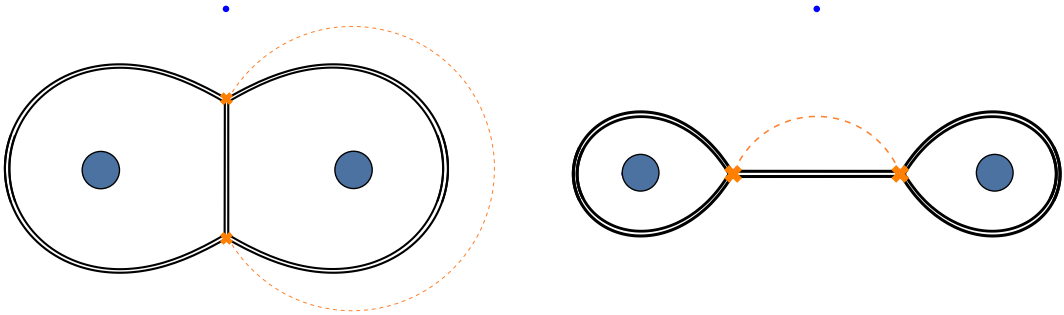}
\caption{This shows the two possible types of Fenchel-Nielsen spectral networks on a single pair of pants where the dotted orange lines are a branch cut.  These spectral networks are called "molecules" in \cite{Hollands:2013qza}.}
\label{fig:mol}
\end{center}
\end{figure}

After $n$ such flips, the interior edges run between the $0^{th}$ exterior vertex preimage to the $n^{th}$ and $(n-1)^{th}$ interior preimage. We can now make sense of the corresponding Darboux coordinates in the limit as $n\to \infty$. First note that as $n\to \infty$ the interior edges approach a parallel line to the interior and exterior edges. This corresponds to a spectral network where there is a single, double wall circling the puncture of $C$ under consideration. If we define $\CY_+$ and $\CY_-$ to be the edges with higher and lower winding number respectively after $n$ flips, then in the $n\to \infty$ limit we can construct the well defined coordinates:
\begin{align}\begin{split}\label{juggcoord1}
\CY_A^{(+)}=\lim_{n\to \infty} \CY_+\CY_-\quad,\qquad
\CY_B^{(+)}&=\lim_{n\to \infty}(\CY_+)^{-n}(\CY_-)^{1-n}~.
\end{split}\end{align}
Similarly, there exists an analogous coordinate system $\{\CY_A^{(-)},\CY_B^{(-)}\}$ for the limit of sending the winding to $-\infty$  which is related by
\be\label{jugcoord2}
\CY_A^{(-)}=(\CY_A^{(+)})^{-1}\quad,\qquad \CY_B^{(-)}=((\CY_A^{(+)})^{1/2}-(\CY_A^{(+)})^{-1/2})^{-4}(\CY_B^{(+)})^{-1}~.
\ee

\subsection{Fenchel-Nielsen Networks}

Now we will discuss a special class of spectral networks called \textit{Fenchel-Nielsen networks} \cite{Hollands:2013qza}. These spectral networks have only double walls corresponding to a set of minimal cuts necessary to decompose the Riemann surface $C$ into a disjoint product of punctured discs and annuli.  This is  a WKB spectral network where $\varphi_2$ is a Jenkins-Strebel differential --- $\varphi_2$ gives a foliation of $C$ by closed paths. Another way of saying this is that a Fenchel-Nielsen spectral network is given by a pants decomposition of $C$ in which on each pair of pants, the spectral network is one of the two networks in Figure \ref{fig:mol}. 

 These spectral networks are referred to as Fenchel-Nielsen-type because the $\fa$-type Fenchel-Nielsen coordinate has a straightforward interpretation in terms of the associated spectral network coordinates.  associated to these networks have a straightforward interpretation as complexified Fenchel-Nielsen coordinates.

Let us take a maximal set of non-intersecting curves $\{\gamma_i\}_{i=1}^{3g-3+n}$ that define a pants decomposition of $C$.
On each pair of pants, there are classes of curves which are homotopic to a subset of the $\{\gamma_i\}$. The holonomy around a curve that is homotopic to such a $\gamma_i$ is given in terms of the spectral network coordinates  \footnote{Note that $\CY_\gamma$ are defined for $\gamma\in H_1(\Sigma;\IZ)$ while $L_\gamma$ is defined for $\gamma\subset C$. Here we use the loose notation where $\CY_\gamma$ for $\gamma\subset C$ is defined as $\CY_{\pi^{-1}(\gamma)|_i}$ the lift under the projection $\pi:\Sigma\to C$ onto one of the sheets. Due to \eqref{eq:branch}, the two choices of lifting are related by inverses and thus are merely a choice of convention. }
\be
\langle L_{\gamma_i}\rangle={\rm Tr}_2\left(\begin{array}{cc} \CY_{\gamma_i}&0\\0&\CY_{\gamma_i}^{-1}\end{array}\right)=\CY_{\gamma_i}+\CY_{\gamma_i}^{-1}~.
\ee
However, we see from before, that this is simply the definition of the Fenchel-Nielsen coordinate $\fa$:
\be\label{fadef}
\CY_{\gamma_i}+\CY_{\gamma_i}^{-1}={\rm Tr}_2 e^\fa
\ee
Wilson line vevs for a fundamental representation of a factor in the (four-dimensional) 
gauge algebra are usually expressed as three-term expressions in the functions 
$\CY_{\gamma}$.  (See e.g.   (10.33) from \cite{Gaiotto:2010be}.)  The relation to the 
above two-term expansion is clarified in equation \eqref{threetermtotwo} above.

\begin{figure}[t]
\centering
\includegraphics[scale=.7]
{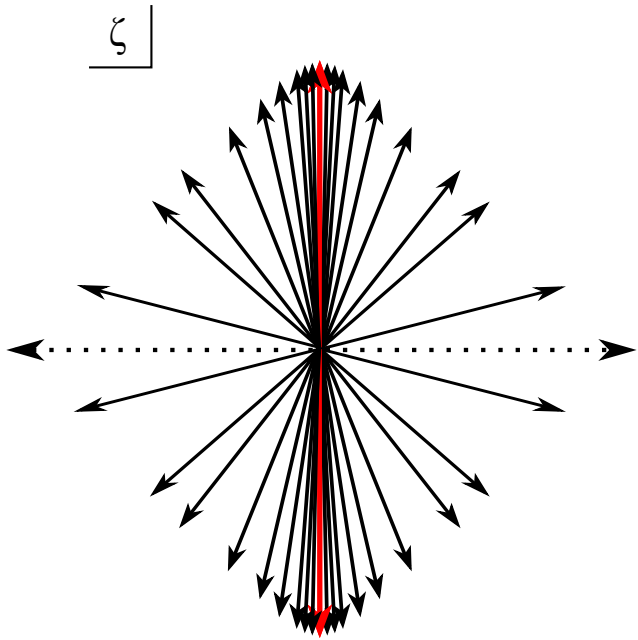}
\caption{This figure shows the structure of the $\CK$-walls in the $\zeta$-plane. 
There are accumulation points (red) on the imaginary axis where the associated WKB spectral network becomes a Fenchel-Nielsen spectral network. }
\label{fig:ZetaPlane}
\end{figure}

In a large class of theories,   such as the ones we study here, Fenchel-Nielsen spectral networks can be obtained from a generic WKB spectral network by performing a juggle.
This requires changing $\zeta$ such that we cross an infinite number of $\CK$-walls.
In the theories we consider, there are infinite number of such walls which accumulate along co-dimension 1 ``accumulation points'' in the $\zeta$-plane. See Figure \ref{fig:ZetaPlane}.
In our setting, sending $\zeta$ to an accumulation point is equivalent to undergoing the infinite number of flips that occur in a juggle, leading to a Fenchel-Nielsen spectral network. See Section \ref{sec:NPFN} for further discussion.


\rmk.~
Recall from the discussion of Section \ref{sec:3}, that the spectral network coordinates 
$\CY_{\gamma}$ is given in \eqref{eq:NPDarboux} and has a semiclassical expansion with an infinite number of non-perturbative corrections.
Since, as we showed above,  we can identify the complexified Fenchel-Nielsen coordinates with spectral network coordinates,
 the Fenchel-Nielsen coordinates $\fa,\fb$ must similarly have an infinite number of non-perturbative corrections to their semiclassical value. We will demonstrate this in the example of the  $SU(2)$ N$_f=0$ theory in Section \ref{sec:SU2Nf0} by computing the leading non-perturbative corrections. 
 




\section{Semiclassical Formulation of BPS States}
\label{sec:SCFBPS}

Because the expectation value of a line defect in a theory of class $\CS$ can be expanded as a  series of Darboux coordinates with coefficients that are framed BPS indices \eqref{eq:Dexp}, the expectation value of the line defect is entirely encoded in the spectrum of framed BPS states. In the semiclassical limit of Lagrangian 4D $\CN=2$ theories, the spectrum of framed BPS states can be described by the index of a Dirac operator on singular monopole moduli space \cite{Brennan:2016znk,Moore:2015szp,Moore:2015qyu}.

The identification of BPS states with the kernel of a Dirac operator arises from the effective description of the dynamics of BPS states in the adiabatic limit via collective coordinates.   The resulting theory is a SQM on (bundles over) singular monopole moduli space with potential \cite{Manton:1981mp,Brennan:2016znk,Moore:2015szp,Moore:2015qyu,Tong:2014yla,Gauntlett:1999vc,Gauntlett:2000ks}. 
Solving for the BPS spectrum is reduced to solving the Dirac equation on singular monopole moduli space coupled to certain bundles over $\CM_{BPS}$.

\subsection{Moduli Space Approximation}

Consider the semiclassical limit of a four-dimensional $\CN=2$ supersymmetric gauge theory with hypermulitiplet matter and 't Hooft line defect insertions where $|X_\infty|>>|Y_\infty|$.
The BPS equations for this theory are given by
\be\label{BPSred}
D_iX=B_i\quad,\qquad D_iY=E_i\quad, \qquad D^i E_i=0~,
\ee
where $\zeta^{-1}\Phi=Y+i X$ is the decomposition of the $\CN=2$ vectormultiplet scalar into real and imaginary parts and $\zeta$ is specified by the 't Hooft defect.
In the background defined by the asymptotic boundary conditions
\begin{align}\begin{split}\label{eq:asbnd}
&X=X_\infty-\frac{\gamma_m}{2r}+...\qquad,\qquad B_i=\frac{\gamma_m}{2 r^2}\hat{r}_i+...\\
&Y=Y_\infty -\frac{g^2}{8\pi r}\gamma_e^\ast+...
\quad,\qquad
E_i=\frac{g^2\gamma_e^\ast}{8\pi r^2}\hat{r}_i+...~,
\end{split}\end{align}
in the limit as $r\to \infty$ with an 't Hooft defect insertion,
the space of solutions of the equations \eqref{BPSred}  is singular monopole moduli space: $\fMM(P,\gamma_m;X_\infty)$.\footnote{Upon choosing a solution $\widehat{A}_a=(A_i,X)\in \fMM$ and specifying boundary conditions \eqref{eq:asbnd}, there is a unique solution of the second equation \eqref{BPSred} for $Y,E_i$. }

We can consistently describe the dynamics of BPS states up to order $O(g^2)$ as dynamics on the moduli space, $\fMM(P,\gamma_m;X_\infty)$. Let us now introduce coordinates $\{z^m\}$ on $\fMM$. These coordinates will be elevated to time dependent fields
\be
z^m:\IR_t\to \fMM~,
\ee
whose variation will describe the dynamics of the BPS states.

Since we are considering the dynamics of BPS states which are $\half$-SUSY, there are still preserved supercharges. This indicates that the $z^m(t)$ have super-partners $\chi^m(t)$ coming from non-trivial fermionic zero modes.

Similarly, there are also fermionic zero-modes $\psi^s(t)$ coming from the hypermultiplet fermions
that couple the theory to the spin-bundle associated with a vector bundle $\CE_{\rm matter}\to \fMM$ of rank \cite{Brennan:2016znk}
\be
rnk_\IR[\CE_{\rm matter}]=\half\sum_{\mu \in \Delta_\CR}n_\rho(\mu)\left\{\langle \mu,\gamma_m\rangle\text{ sign}(\langle \mu,X_\infty\rangle+m_x) +\sum_j|\langle \mu,P_j\rangle|\right\}~,
\ee
where $m_x={\rm Im}[\zeta^{-1}m_f]$.
Thus, the dynamics of the full effective SQM couples to the bundle
\be
Spin(\CE_{\rm matter})\otimes S\fMM\longrightarrow \fMM~,
\ee
which is given in terms of the collective coordinates $\{z^m(t)$, $\chi^m(t)$, $\psi^a(t)\}$.

The corresponding supercharges are explicitly of the form
\begin{align}\begin{split}
\hat{Q}^a=\chi^m (\tilde{\IJ}^a)_{m}^{~n}(\dot{z}_n-G(\CY_\infty)_n)~,
\end{split}\end{align}
where
\be
\IJ^a=(\CJ^r,\mathds{1})\quad,\qquad \tilde{\IJ}^a=(-\CJ^r,\mathds{1})~.
\ee
and $\{\CJ^r\}$ are the complex structures on $T\fMM$ and $G(\CY_\infty)_m$ is a triholomorphic vector field that generates a global gauge transformation \footnote{ More generally, the vector fields $G(H_I)_m$, which generate global gauge transformations along $H_I\in \ft$, are defined by
\begin{align}\begin{split}
\hat{D}^2 \epsilon_{H_I}=0\quad,\qquad \lim_{|\vec{x}|\to \infty} \epsilon_{H_I}=H_I\quad,\qquad \hat{D}_a\epsilon_{H_I}=-G(H_I)^m\delta_m \widehat{A}_a~,
\end{split}\end{align}
where $\hat{D}_a$ is the covariant derivative with connection $\widehat{A}_a\in \fMM(P,\gamma_m;X_\infty)$ and the $\{\delta_m \widehat{A}_a\}$ form a basis of the tangent space $T_{[\widehat{A}_a]}\fMM(P,\gamma_m;X_\infty)$. } along $\CY_\infty\in \ft$ where
\be
\CY_\infty={\rm Re}[\zeta^{-1}a_D]=\frac{4\pi}{g^2}Y_\infty+\frac{\vartheta}{2\pi}X_\infty~,
\ee
where $\zeta$ is defined by the line defect.

In the Hamiltonian formulation, the supercharge operator becomes a twisted Dirac operator:
\be
\hat{Q}^a=
i\gamma^m(\tilde\IJ^a)_m^{~n}(\CD_n- iG(\CY_\infty)_n)~,
\ee
where $\CD_n$ is the spin coviariant derivative on $Spin(\CE_{\rm matter})\otimes S\fMM\to \fMM$. Thus, BPS states are those that are in the kernel of the supercharge Dirac operators.

Now from the supersymmetry algebra
\be
\{\hat{Q}^a,\hat{Q}^b\}=2\delta^{ab}(\hat{H}+\text{Re}[\zeta^{-1}Z])~,
\ee
we see that if a BPS state is in the kernel of \textit{any} of the supercharges, then it is in the kernel of \textit{all} of the supercharges. Therefore, the spectrum of BPS bound states can be determined by considering the kernel of the $4^{th}$ supercharge
\be
\hat{Q}^4=i \gamma^m (\CD_m-i G(\CY_\infty)_m)\equiv i\slashed{D}^{\CY}~.
\ee
Additionally, since the electric charge and flavor charge operators commute with the Hamiltonian, and hence the supercharge operator, we can simultaneously diagonal the two. This leads to a decomposition of the Hilbert space of BPS states into $\gamma=\gamma_m\oplus\gamma_e\oplus\gamma_f$-isotypical
\be
\CH^{BPS}_{L_{p,0}}=\bigoplus_{\gamma\in \Gamma}\CH^{BPS}_{L_{p,0},\gamma}~.
\ee
Therefore, we can identify
\be
\text{dim}[\CH^{BPS}_{L_{p,0},\gamma}]={\rm Ind}_{L^2}\Big[\slashed{D}^{\CY}\Big]^{\gamma_{e\oplus\gamma_f}}_{\fMM}=\fro\left(\gamma;u\right)\quad, \quad \gamma=\gamma_m\oplus\gamma_e\oplus\gamma_f~,
\ee
where here we mean the $\gamma_{e}\oplus\gamma_f$-isotypical component of the $L^2$ index of the Dirac operator $\slashed{D}^{\CY}$ on $\fMM(P,\gamma_m;X_\infty)$.

\subsection{Fenchel-Nielsen Spectral Networks and The Semiclassical Region}
\label{sec:NPFN}

Now we can connect the formalism of semiclassical BPS states and spectral networks. Recall that  the the expectation value of line defects is determined by the framed BPS index as in \eqref{eq:Dexp}. Therefore,  the index of the Dirac operator from the previous section can be used to determine the expectation value of an 't Hooft defect in the semiclassical limit.

Since we want to compare  to the localization computation, which is naturally expressed in terms of Fenchel-Nielsen coordinates, one would hope to use Fenchel-Nielsen spectral networks and the associated Dirac operators.  In order to implement this we need to know: 1.) if there exists Fenchel-Nielsen spectral networks in the semiclassical limit and 2.) where in parameter space these spectral networks exist so that we can compare to indices of Dirac operators. In  this section we will show that such spectral networks exist in the semiclassical limit, but that they only exist in parameter space where the moduli space approximation breaks down.



The question of whether or not a Fenchel-Nielsen network exists is equivalent to the question of whether or not there exists a Jenkins-Strebel differential on $C$ that encodes the data of the theory in some semiclassical limit. The data of the differential is $(u,\zeta,m)\in \CB\times U(1)\times \ft_F$.

The existence of Jenkins-Strebel differentials on a Riemann surface $C$ with punctures are studied by Liu \cite{Liu,Liu0}. There, Liu shows  that given a decomposition of $C$ into a collection of punctured disks $\{D_m\}$ and annuli $\{R_k\}$, there exists a uniquely determined real Jenkins-Strebel differential with closed trajectory $\varphi_2$ with fixed monodromy $m_i\in \IR$ around each puncture and height $h_k\in \IR$ around each annuli where the height is defined as
\be
h_k={\rm Inf}_{\gamma_k}\oint_{\gamma_k } \left|{\rm Im}\sqrt{\varphi_2}\right|~,
\ee
where the infimum is taken over all paths that run between the boundaries of $R_k$.
Note that the Fenchel-Nielsen spectral network is exactly given by the union of the boundaries of these component disks and annuli.



\begin{figure}[t]
\centering
\includegraphics[scale=0.7,clip,trim=2cm 17cm 9cm 3cm]{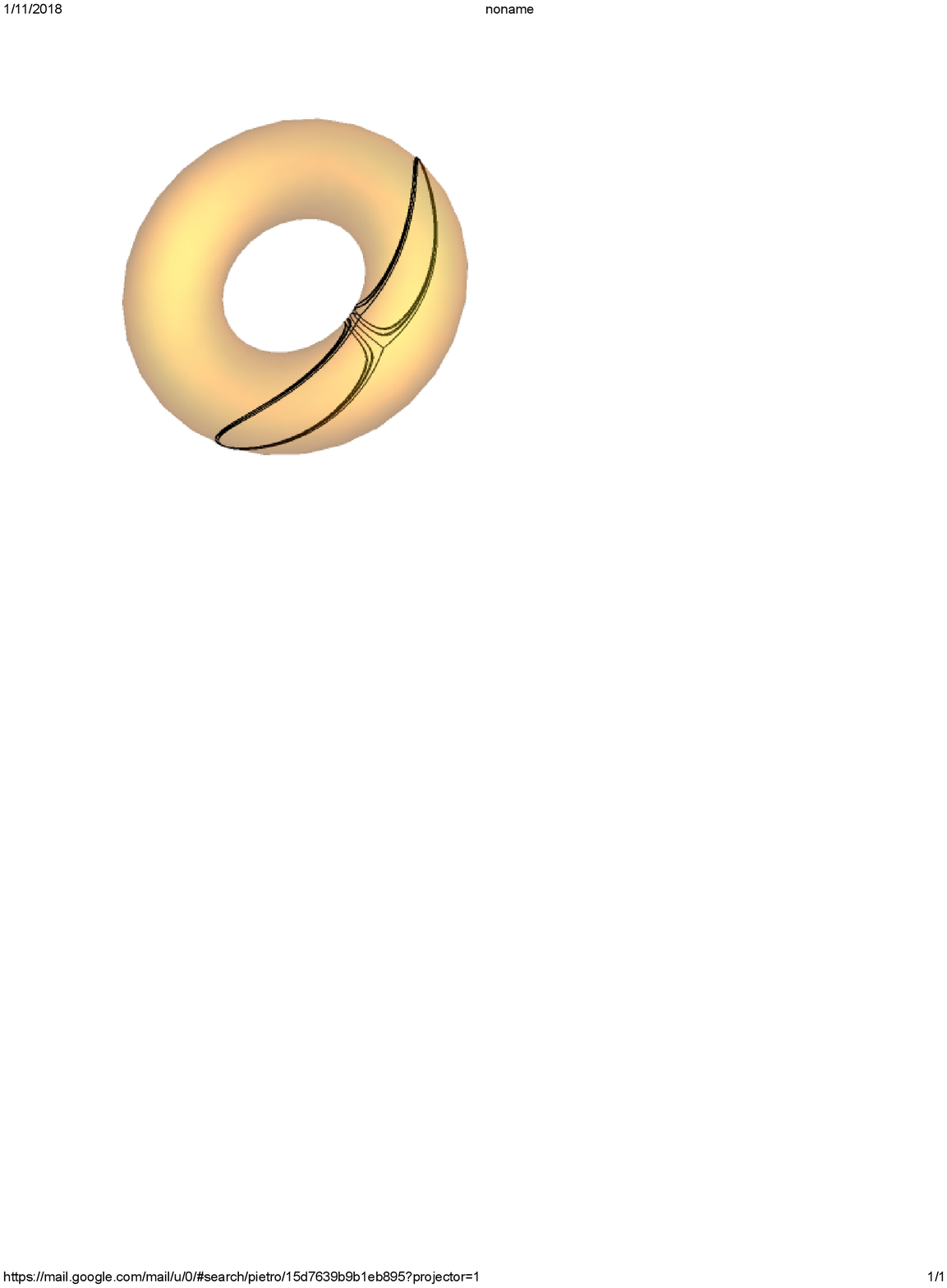}
\includegraphics[scale=0.7,clip,trim=2cm 17cm 9cm 3cm]{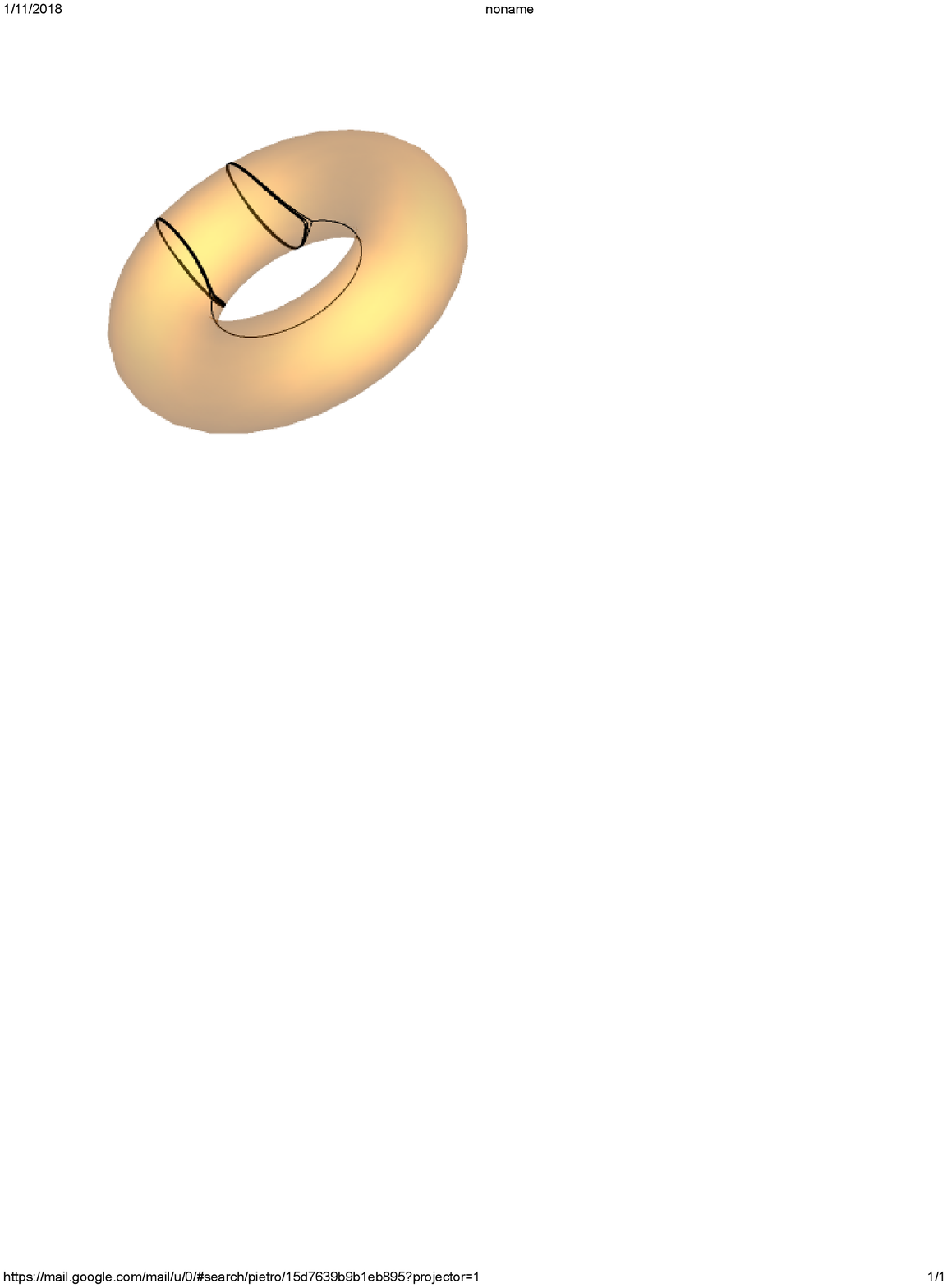}
\caption{This figure shows the explicit decomposition of the UV curve $C=T^2\slash \{0\}$ into disks and annuli for the 4D $SU(2)$ $\CN=2^\ast$ theory in two different ways. Note that the boundary of these components give rise to the Fenchel-Nielsen spectral networks corresponding to both types of fundamental molecules. Which type of Fenchel-Nielsen molecules appear in the spectral network is dependent on the relative holonomies of the cuts. See \cite{Hollands:2013qza} for details.}
\label{fig:FNTorus}
\end{figure}

Now consider as an example the case of the 4D $SU(2)$ $\CN=2^\ast$ theory. 
This
theory is constructed as a theory of class $\CS$ by taking $C$ to be a torus with a single puncture. 
This theory comes with a complex 2-dimensional parameter space defined by $u\in \CB\cong \IC$ and the complex mass parameter of the hypermultiplet. $C$ can be decomposed as an annulus $R_a$ and a punctured disk $D_{m}$.  See Figure \ref{fig:FNTorus} for the example of the 4D $SU(2)$ $\CN=2^\ast$ theory where $C=T^2\slash \{0\}$.\footnote{We would especially like to thank Pietro Longhi for providing these figures.} Thus, there is a $3$ dimensional family (specifying $m$, $\gamma$, and $\zeta$) of Jenkins-Strebel differentials which forms a real co-dimension $1$ subspace of parameter space. This suggests that there could exist a Jenkins-Strebel differential in the semiclassical limit  ($|u|\to \infty$ ) and therefore that there could exist a Fenchel-Nielsen spectral networks in the semiclassical limit. This has been confirmed by numerical computations.\footnote{We would like to thank Pietro Longhi for sharing his numerical computation for the $SU(2)$ $\CN=2^\ast$ theory and for making the authors aware of Liu's work on Jenkins-Strebel differentials. }

Now recall that for  a WKB Fenchel-Nielsen spectral network, the \emph{real} Jenkins-Strebel differential is related to the Seiberg-Witten differential as
\be
\varphi_2=\zeta^{-2}\lambda^2_{SW}~.
\ee
Asking that $\varphi_2$ as defined by this equation is a Jenkins-Strebel differential defines the Fenchel-Nielsen locus in $\CB^\ast \times \IC^\ast$.

As usual in Seiberg-Witten theory, the periods of $\lambda_{SW}$  give the vev's of the Higgs field and mass parameters. In our case the UV curve is given by $C=T^2\backslash\{0\}$. This means that if we pick a basis of $H_1(\bar{\Sigma};\IZ)=\text{span}_\IZ\{A,B\}$,
\be
\oint_A \lambda_{SW}=a\quad,\qquad \oint_B\lambda_{SW}=a_D\quad,\qquad \oint_{D_{p}}\lambda_{SW}=m_f~,
\ee
where $D_{p}$ is a loop circling the puncture and $m_f$ is the mass of the adjoint hypermultiplet. In this notation, the condition that $\varphi_2$ is a Jenkins-Strebel differential (and hence gives rise to a Fenchel-Nielsen-type WKB spectral network) is that
\be
\oint_A\zeta^{-1}\lambda_{SW}\in \IR\quad,\qquad \oint_{D_{m_i}}\zeta^{-1} \lambda_{SW}\in \IR~,
\ee
which can be rewritten as
\be
\text{Im}[\zeta^{-1}a]=X_\infty=0\quad,\qquad \text{Im}[\zeta^{-1}m_f]=m_x=0~.
\ee
This locus in parameter space, which we will call the \emph{Fenchel-Nielsen locus}, is an accumulation point of $\CK$-walls in the $\zeta$-plane and we will denote the associate phase in $U(1)$ as $\zeta_{FN}$.

Unfortunately, the Fenchel-Nielsen locus is exactly where the moduli space approximation, which gives the identification between the framed BPS index and the index of a Dirac operator on singular monopole moduli space, breaks down. In the limit $X_\infty\to 0$, the space $\fMM(P_n,\gamma_m;X_\infty)$ (and $\CM(\gamma_m;X_\infty)$) are not defined. The reason is that the semiclassical expression for the central charge is given by
\begin{align}\begin{split}
\zeta^{-1}Z_{\gamma}&=
-\left[\frac{4\pi}{g^2}(\gamma_m,X_\infty)-\langle \gamma_e
,Y_\infty\rangle\right]+i\left[\frac{4\pi}{g^2}(\gamma_m,Y_\infty)+\langle \gamma_e
,X_\infty\rangle\right]~.
\end{split}\end{align}
Thus the BPS mass $M_{BPS}={\rm Re}[\zeta^{-1}Z_{\gamma}]$ for a monopole goes to zero as we scan $\zeta$ such that $X_\infty \to 0$. However, we know that monopoles do not become massless in the semiclassical limit. Thus, 
we can deduce that the non-perturbative quantum effects must become large and therefore the effective SQM description above must break down.

However, by taking $|X_\infty|,|Y_\infty|\to \infty$ as $|X_\infty|/|Y_\infty|\to 0$, we can still identify framed BPS indices with the index of a Dirac operator for phases which are arbitrarily close to the Fenchel-Nielsen locus. This will allow us to give an index theorem for the supercharge Dirac operators almost everywhere on the $\zeta$-plane. See Figure 1 of \cite{Moore:2015qyu} or Figure 4 of \cite{Moore:2015szp} for more details.

 The above analysis makes it clear that there always exists $SU(2)$ Fenchel-Nielsen networks (and in fact all $SU(N)$-type Fenchel-Nielsen spectral networks) in the semiclassical limit. These exist on the locus where all of the masses and $a_i=\oint_{A_i}\lambda_{SW}$ have the same phase. Such a spectral network can be constructed by gluing together pairs of pants with semiclassical Fenchel-Nielsen spectral networks on them by the procedure of \cite{Hollands:2013qza}. The only condition here is that the Fenchel-Nielsen spectral networks all have the associated phase.

\rmk.~
Recall that a Fenchel-Nielsen spectral network corresponds to a WKB spectral network with a Jenkins-Strebel differential. This is defined by decomposing the Riemann surface $C$ into a collection of annuli and punctured disks. On each component, the flow lines of $\varphi$ give a foliation of curves that are homotopic to the boundary components. If we consider infinitesimally deforming the phase $\zeta$ associated to the quadratic differential, we find that the flow lines on each component are no longer homotopic to the boundary components, but rather spiral into them with a very large winding number. Thus, as we send $\zeta\to \zeta_{FN}$ the flow lines of $\varphi_2$ twist around the boundary components infinitely many times until they form closed paths, producing a Fenchel-Nielsen spectral network. This infinite spiraling indicates that Fenchel-Nielsen spectral networks can be achieved by performing a juggle on a WKB spectral network where all  physical parameters have aligned phases. Using the procedure from Section 8.4 of \cite{Hollands:2013qza}, one can identify the limiting coordinates \eqref{juggcoord1} with the Fenchel-Nielsen coordinates $\CY_A^{(+)}=e^{\fa}$, 
while $\CY_B^{(+)}$ defines a choice of $e^{\fb}$. \footnote{Note that we could also approach the Fenchel-Nielsen locus in the opposite direction. The procedure from \cite{Hollands:2013qza} in conjunction with the relation between the two limiting coordinates \eqref{jugcoord2}, correspond to two different choices of Fenchel-Nielsen coordinates. 
Equation (7.52) of \cite{Gaiotto:2009hg} shows that $\fa$ is well-defined and $\CY_B^\pm$  define two choices of $\fb$ that are related by \eqref{jugcoord2}.
}

 Thus, the Darboux coordinates associated to Fenchel-Nielsen spectral networks in the cases we are studying can be obtained by acting on a generic set of spectral network coordinates by an infinite number of cluster coordinate transformations. The resulting spectral network coordinates are those that result from the flip \eqref{juggcoord1}. This gives a recursion formula for the Darboux coordinates that can be ``integrated" to give a relation between the Darboux coordinates of a spectral network in any chamber and the Fenchel-Nielsen coordinates which are used in localization computations. This will be the primary computational tool that we will use to construct an index theorem and give a formula for the characteristic numbers in the next section.

%

\section{Index Theorem and Characteristic Numbers}
\label{sec:sec5}

In this section we will compare the different methods of computing the expectation value of 't Hooft defects in 4D  $\CN=2$ $G=SU(2)$ asymptotically free theories with fundamental and hypermultiplet matter. We will outline how this comparison can be used to give an index theorem for Dirac operators on singular monopole moduli spaces and give the characteristic numbers of certain Kronheimer-Nakajima spaces\footnote{These are the transversal slice to the stratum of the bubbling locus of singlar monopole moduli spaces. See \cite{Brennan:2018yuj,Nakajima:2016guo} for  details.}. We will explicitly show these for the $SU(2)$ ${\rm N}_f=0$ theory.

\subsection{General Theory}

Consider an $\CN=2$ $SU(2)$ Lagrangian theory of class $\CS$ with mass parameters of identical phase. Now pick a point in the semiclassical limit of the Coulomb branch away from the Fenchel-Nielsen locus. We are interested in computing the expectation values of a 't Hooft defect which is specified by an integer $p$ and a phase $\zeta$.

Now consider comparing the localization and spectral network result for the expectation value of 't Hooft defects.
Localization requires introducing an IR regulating $\half\Omega$-deformation and expresses the  expectation value in terms of Fenchel-Nielsen coordinates. This coordinate expansion is well defined almost everywhere in a simply connected region on the Coulomb branch  and is independent of the phase of $\zeta$ there due to trivial monodromy. The spectral networks computation however, is not independent of the phase $\zeta$. Rather, it is different in each chamber $c_n\subset \IC_\zeta$ of the $\zeta$-plane.\footnote{Here $\zeta$ changes the decomposition of $u$ into $X_\infty,Y_\infty$.} The reason is that the spectral network undergoes topology change at each $\CK$-wall and
 hence has a different set of associated Darboux coordinates 
  in each chamber $c_n\subset \IC_\zeta$.


Away from the Fenchel-Nielsen locus, the spectral network coordinates are not Fenchel-Nielsen coordinates, but rather are Darboux coordinates which are related to the localization Fenchel-Nielsen coordinates by an infinite sequence of Kontsevich-Soibelman transformations. 

Due to the ``simple" transformation properties of the spectral network coordinates, these coordinate transformations can be integrated to determine the mapping between Fenchel-Nielsen coordinates and the Darboux coordinates in every chamber. This can be achieved as follows. First, solve for the expectation value of the minimal Wilson and 't Hooft defects  in a generic WKB spectral network  of choice. We will assign the chamber in the $\zeta$-plane in which we have computed these as the $c_0$ chamber.  Now by tuning the phase of $\zeta$, we will cross walls of marginal stability which takes us from the $c_n$ chamber to the $c_{n\pm1}$ chamber depending on the direction we tune $\zeta$.

Now we can solve for the expectation values in all chambers by solving the recursive $\CK$-wall crossing formulas \cite{Gaiotto:2010be}:
\be
\langle L_{1,0}\rangle_{\zeta\in c_n}(\CY_{\gamma_i})=\langle L_{1,0}\rangle_{\zeta\in c_{n-1}}(\CK_{\gamma_{n}}\cdot \CY_{\gamma_i})\quad, \quad \langle L_{0,1}\rangle_{\zeta\in c_n}(\CY_{\gamma_i})=\langle L_{0,1}\rangle_{\zeta\in c_{n-1}}(\CK_{\gamma_{n}}\cdot \CY_{\gamma_i})~,
\ee
where $\langle L\rangle_{\zeta\in c_n}$ is the expectation value of $L$ computed using the WKB spectral network acssociated to $\zeta\in c_n$ and the $\CK$-wall between the $c_n$ and $c_{n-1}$ chamber is $\widehat{W}(\gamma_n)$.

After we set the $\half \Omega$ deformation parameter $\epsilon_+\to 0$, we can then compare the localization expression of the expectation value of the Wilson and `t Hooft defects to their expression in terms of the spectral network coordinates in a generic chamber chamber $c_i$. Inverting these formulas allows us to solve for $(\fa,\fb)$ in terms of the $\CY_\gamma$ in some fixed chamber $c_i$.

Then, by combining this with the solution with the KS-wall crossing formulas, we then have an expression for the Fenchel-Nielsen coordinates $(\fa,\fb)$ in terms of the $\CY_{\gamma}$ in all chambers $c_n$. Inverting the formulas, one obtains an (admittedly complicated) expression for the $\fa, \fb$ in terms of the $\CY_{\gamma}$.

We can then take these expressions and substitute the expression for $\fa,\fb$ in terms of the $\CY_\gamma$ into the localization expression above.  By identifying the coefficients of the Laurent expansion with that of the spectral network computation we arrive at an expression for the framed BPS indices in every chamber. Then by using the relation of the index of the Dirac operator to the framed BPS indices in the semiclassical limit away from the Fenchel-Nielsen locus
\begin{align}\begin{split}
&\sum_{\gamma\in \Gamma} {\rm Ind}\Big[\slashed{D}^{\CY}\Big]^{\gamma_e\oplus\gamma_f}_{\fMM(P,\gamma_m;X_\infty)}\CY_\gamma=\sum_{\gamma\in \Gamma}\fro\left(\gamma;L_{[P,0]},c_n\right)\CY_\gamma\\
&=\left\{\sum_{|\vv|\leq |P|}e^{ (\vv,\fb)}\big(F(\fa)\big)^{|\vv|} \left[\lim_{\xi\to0}\int_{\widetilde\CM^\xi_{KN}(P,\vv)}e^{\omega+\mu_T}\widehat{A}_{T}(T\widetilde\CM_{KN})\cdot C_{T\times T_F}(\CV(\CR))\right]\right\}_{\substack{e^{\fa}=f_\fa(\CY_\gamma)\\e^{\fb}=f_\fb(\CY_\gamma)}}~,
\end{split}\end{align}
we get an index formula for the associated Dirac operator in all chambers arbitrarily close to the Fenchel-Nielsen locus.
%

Similarly, we can substitute the expression for the Darboux coordinates $\CY_\gamma$ in the $c_n$ chamber in terms of the Fenchel-Nielsen coordinates $\fa,\fb$ into the spectral network computation. Then, identifying the coefficients of the Laurent expansion in terms of the exponentiated Fenchel-Nielsen coordinates on both sides
\begin{align}\begin{split}
\sum_{|\vv|\leq P}e^{ \fb}\big(F(\fa)\big)^{|\vv|}\,Z_{mono}(\fa,m;P,\vv)=\sum_{\gamma\in \Gamma} \fro(\gamma;L_{[P,0]},c_n)\CY_\gamma(\fa,\fb)~, \\
Z_{mono}(\fa,m;P,\vv)=\lim_{\xi\to0}\int_{\widetilde\CM^\xi_{KN}(P,\vv)}e^{\omega+\mu_T}\widehat{A}_{T}(T\widetilde\CM_{KN})\cdot C_{T\times T_F}(\CV(\CR))~,
\end{split}\end{align}
allows us to express the characteristic numbers that determine $Z_{mono}(\fa,m;P,v)$ as a rational function of exponentiated Fenchel-Nielsen $\fa$-coordinates, masses, and framed BPS indices.

\rmk~ Note that there is an additional subtlety in the case of the $\CN=2$ $SU(2)$ ${\rm N}_f=4$ theory. The reason is that $Z_{mono}(\fa,m;P,v)$ is not entirely given by a characteristic number but rather has an additional contribution from states on the Coulomb branch of an associated SQM \cite{Brennan:2018rcn}.

\subsection{Example: $SU(2)$ ${\rm N}_f=0$ Theory}
\label{sec:SU2Nf0}

Now we will apply the above, discussion to determine the framed BPS indices for the $SU(2)$ $ {\rm N}_f=0$ theory. This will produce an index-like formula for a Dirac operator coupled to a hyperholomorphic vector field $G_n(\CY_\infty)$ on singular monopole moduli space.

The expectation value of the 't Hooft defect in the N$_f=0$ theory is given in terms of Fenchel-Nielsen coordinates as
\be
\langle L_{p,0}\rangle_{Loc}=\sum_{0\leq \vv\leq p} \cosh (\vv,\fb) \big(F(\fa)\big)^{\vv} Z_{mono}(\fa;P,\vv)~,
\ee
where
\be
P={\rm diag}(p, -p)\quad, \quad \vv={\rm diag}(v,-v)~.
\ee
In the case where $\epsilon_+=0$, which is necessary for comparing with the Fenchel-Nielsen and Dirac operator expressions, these have the simple form
\begin{align}\begin{split}
&\langle L_{0,1}\rangle_{Loc} =e^{   \fa}+e^{-   \fa}\quad,\qquad \langle L_{1,0}\rangle_{Loc}=
\frac{e^{ \fb}+e^{-\fb}}{2\sinh( \fa)}~.
\end{split}\end{align}

As shown in \cite{Gaiotto:2010be,Moore:2015szp}, the expectation value of the 't Hooft defect of minimal charge in terms of Darboux coordinates in the chamber $c_n$ is given by \footnote{Recall that $L_{1,0}$ is the minimal 't Hooft defect and $L_{0,1}$ is the minimal Wilson defect.}
\be\label{SYMLineDef}
\langle L_{1,0}\rangle_{\zeta\in c_n}=
\frac{1}{\CX_m \CX_e^n}\left(U_n(f_n)-\frac{1}{\CX_e}U_{n-1}(f_n)\right)
\quad, \quad \langle L_{0,1}\rangle_{\zeta\in c_n}=2 f_n~,
\ee
where
\be\label{fnWilson}
f_n=\half\left(\CX_e+\frac{1}{\CX_e}\Big(1+\CX_m^2 \CX_e^{2n+2}\Big)\right)~,
\ee
and
\be
\CX_m=\CY_{\half H_\alpha}\quad,\quad \CX_e=\CY_{\half \alpha}~.
\ee
Here we take $n\in \IZ_+$ to denote the chamber $c_n$ in the $\zeta$-plane and the notation $U_n$ to denote the Tchebyshev polynomial of the second kind:
\be
U_{n-1}(\cos(x))=\frac{\sin(nx)}{\sin(x)}~.
\ee
In this theory, the Fenchel-Nielsen locus is given by $\langle \alpha,X_\infty\rangle=0$. \footnote{Note, that here we have defined $X_\infty\in \ft$ to lie in the positive chamber. We will thus assume that $\langle \alpha,X_\infty\rangle\geq0$. }
Let us pick a $u\in \CB$ such that $\Phi_\infty=\pm iX_\infty$ when $\zeta\in \IR$ and $\Phi_\infty=\pm Y_\infty $ when $\zeta \in i\IR$. We can now identify the imaginary axis as the Fenchel-Nielsen locus.

The spectrum of the vanilla BPS states in the semiclassical region are given by 
\be
\gamma=\pm \alpha\quad, \quad \gamma_n^\pm=\pm  H_\alpha\oplus n\, \alpha\quad,~n\in \IZ~,
\ee
with BPS indices
\be\label{SYMBPSInd}
\Omega(\gamma;u)=\begin{cases}
-2&\gamma=\pm\alpha\\
1&\gamma=\gamma_n^\pm\\
0&else
\end{cases}
\ee
%
Thus, the phase of the central charge corresponding to a state with charge $\gamma=\gamma_m\oplus\gamma_e$ is given by
\begin{align}\begin{split}
{\rm phase}(Z_{\gamma})=-\arctan\left[\frac{(\gamma_m,Y_\infty)}{(\gamma_m,X_\infty)}+\frac{g^2}{4\pi}\left(\frac{\langle \gamma_e,X_\infty\rangle}{(\gamma_m,X_\infty)}+\frac{(\gamma_m,Y_\infty)\langle \gamma_e,Y_\infty\rangle}{(\gamma_m,X_\infty)^2}\right)\right]+O(g^4)~,\\
=-\arctan\left\{\frac{(\gamma_m,Y_\infty)}{(\gamma_m,X_\infty)}\left[1+\frac{g^2}{4\pi}\left(\frac{\langle \gamma_e,Y_\infty\rangle}{(\gamma_m,X_\infty)}+\frac{\langle\gamma_e,X_\infty\rangle }{(\gamma_m,Y_\infty)}\right)\right]\right\}+O(g^4)~.
\end{split}\end{align}

 Without loss of generality, we can restrict to the case $\zeta$ in the positive real half-plane (the other cases follow analogously). We are now only concerned with the phase of BPS states whose $\CK$-walls are in the positive real half-$\zeta$ plane. These BPS states have charges $\gamma_n^-$ with $\CK$-walls along the phases
\be
{\rm phase}(Z_{\gamma_n^-})=-\arctan\left\{\frac{(H_\alpha,Y_\infty)}{(H_\alpha,X_\infty)}\left[1-\frac{n\, g^2}{4\pi}\left(\frac{(H_\alpha,Y_\infty)}{(H_\alpha, X_\infty)}+\frac{(H_\alpha,X_\infty)}{(H_\alpha,Y_\infty)}\right)\right]\right\}~,
\ee
to order $O(g^4)$. Note that the phases of the chentral charges are ordered
\be
{\rm phase}(Z_{\gamma_n^-})>{\rm phase}(Z_{\gamma_{n-1}^-})~,
\ee
in the positive real half-plane. 
We can now define the chambers $c_n\subset \widehat\IC^\ast$ as
\begin{align}\begin{split}
c_n:=\Big\{\zeta \in \widehat\IC^\ast~|~{\rm phase}(Z_{\gamma_n})>{\rm phase}(\zeta)>{\rm phase}(Z_{\gamma_{n-1}})\Big\}~.
\end{split}\end{align}
At the $\CK$-wall defined by $\widehat{W}(\gamma_n^-)$, the $X_{i,n-1}$ mutate as:
\be
\CK_{\gamma_n^-}X_{1,n-1}=(1+X_{1,n-1} X_{2,n-1}^n)^{-2n}X_{1,n-1}\quad, \quad \CK_{\gamma_n^-}X_{2,n-1}=(1+X_{1,n-1}X_{2,n-1}^n)^2X_{2,n-1}~.
\ee

By comparing with  the computation of $\langle L_{1,0}\rangle$ and $\langle L_{0,1}\rangle$ using localization, we can determine the coordinate transformation relating the Fenchel-Nielsen coordinates to the spectral network coordinates in the $c_n$ chamber:
\begin{align}\begin{split}\label{eq:DarbouxtoFN}
&\CX_m=-\frac{b(a^2-1)(1+a^{2n+2}b^2)^n}{a(1+a^{2n}b^2)^{n+1}}\quad,\qquad \CX_e=\frac{a(1+a^{2n}b^2)}{(1+a^{2n+2}b^2)}~,
\end{split}\end{align}
where
\be
a=e^{\fa}\qquad, \qquad b=e^{ \fb}~.
\ee
We can now invert the expressions \eqref{eq:DarbouxtoFN} to get
\begin{align}\begin{split}\label{FAtoSN}
&a=f_n-\sqrt{f_n^2-1}\quad,\qquad
b=\frac{\sqrt{a-\CX_e}}{\sqrt{a^{2n+1}(a \CX_e-1)}}~,\\
& f_n=\half\left(\CX_e+\frac{1}{\CX_e}+\CX_m^2 \CX_e^{2n+1}\right)~.
\end{split}\end{align}
Note that both of these pairs of expressions requires matching the semiclassical expressions for $X_{1,0},X_{2,0},e^\fa,e^\fb$.

This can be used to show explicitly  that the spectral network coordinates approach the Fenchel-Nielsen coordinates as we approach the Fenchel-Nielsen locus.
Sending the phase $\zeta\to \zeta_{FN}$ can then be achieved by sending $n\to \infty$ or $n\to -\infty$. Going to the Fenchel-Nielsen locus rotates the phase of $\Phi_\infty=\zeta(Y_\infty+i X_\infty)$ so that, in the limit $n\to \pm\infty$, $\pm \langle \alpha,Y_\infty\rangle>0$. Then from the expression for $\CY_\gamma$  \eqref{eq:NPDarboux}, we see that $\lim_{n\to \pm \infty} \CY_{\gamma_n}$ is exponentially suppressed
\be
\lim_{n\to \pm\infty} \CY_{\gamma_n}\sim\lim_{n\to \infty} e^{-(|n|+1)\pi R |\langle \alpha,Y_\infty\rangle|}
\times O(e^{-4\pi^2 R/g^2})=0~.
\ee
Thus, sending $\zeta\to \zeta_{FN}$ reduces the standard three-term expansion of the value of the Wilson line to
\be\label{threetermtotwo}
\lim_{n\to \pm \infty}\langle L_{1,0}\rangle_n=\CY_{\half\alpha}+\CY_{-\half \alpha}~.
\ee
This allows us to identify $\CY_{\half\alpha}=e^{\pm\fa}$. \footnote{As we will see below, the signs are correlated with the different limits $n\to \pm \infty$.}

The coordinate identification above can also be seen from taking limits of the coordinate transformation expressions \eqref{eq:DarbouxtoFN}. As $\zeta\to \zeta_{FN}$, we pass through chambers with $n\to \pm\infty$. From the semiclassical expression for $\fa$ in \eqref{absemi}, we see that in this region 
\be
\lim_{n\to \pm \infty}\left|e^{\pm \langle \alpha,\fa\rangle}\right|_{\zeta\in c_n}<1~.
\ee
Therefore:
\be
\lim_{n\to \pm \infty}\CX_e=\lim_{n\to \pm \infty}\CY_{\half\alpha}=e^{\pm\fa}\big{|}_{\zeta=\zeta_{FN}}~.
\ee
Similarly we can apply this method to $\CX_m=\CY_{\half H_\alpha}$ to find
\be
\lim_{n\to \pm \infty}\CY_{\half H_\alpha}=-\sinh(\fa)e^{\pm \fb}\big{|}_{\zeta=\zeta_{FN}}~.
\ee

Using the identification of the Fenchel-Nielsen coordinates with spectral network coordinates in $c_n$ \eqref{FAtoSN}, it is possible to compute explicitly the non-perturbative corrections to $\fa,\fb$. 
From the results of \cite{Gaiotto:2008cd}, we know that the corrections to the semiclassical contribution of $\CY_\gamma$ are given by solving the recursive formula
\begin{align}\begin{split}
\log\,\CY_\gamma(u,\theta,\zeta)&=\log\, \CY_\gamma^{sf}(u,\theta,\zeta)\\
&+\sum_{\gamma'}\Omega(\gamma';u)\frac{\langle \gamma',\gamma\rangle}{4\pi i }\int_{\ell_{\gamma'}}\frac{d\zeta'}{\zeta'}\frac{\zeta'+\zeta}{\zeta'-\zeta}\log(1-\CY_{\gamma'}(u,\theta,\zeta'))~,
\end{split}\end{align}
where $\CY_\gamma^{sf}$ is the semi-flat term which is the semiclassical expression in \eqref{eq:NPDarboux} and $\ell_{\gamma'}$ is the ray in the $\zeta$-plane along the $\CK$-wall $\CK_{\gamma'}$. 

To compute the first order non-perturbative corrections to $\fa,\fb$, we need to use the fact that the semi-flat expressions for $\CX_e,\CX_m$ are of order $\CX_e^{sf}\sim O(1)$ while $\CX_m^{sf}\sim O\left(e^{-\frac{4\pi}{g^2}}\right)$. Using this, we can expand the 
expressions for $\fa,\fb$ as a Laurent series in $\CX_m$: 
\begin{align}\begin{split}
e^\fa&=\CX_e\left[1+\frac{\CX_e^{2n+1}}{\CX_e-1/\CX_e}\CX_m^2+O(\CX_m^4)\right]~,\\
e^\fb&=-\frac{\CX_m}{\CX_e-\CX_e^{-1}}\left[1-\frac{\CX_e^{2n+1}\CX_m^2}{(\CX_e-1/\CX_e)^3}f_n(\CX_e)+O(\CX_m^4)\right]~,
\end{split}\end{align}
where 
\be
f_n(\CX_e)=\CX_e+n (\CX_e-1/\CX_e)~.
\ee
Then by taking into account the non-perturbative corrections to $\CX_e,\CX_m$ and comparing orders in $e^{-\frac{4\pi}{g^2}}$, we can compute the first order non-perturbative corrections to the leading (semiclassical) expression for $\fa,\fb$ as 
\begin{align}\begin{split}
\fa_{n.p.}^{(1)}&=\log[\CX_e^{n.p.(1)}]+\log\left[1+\frac{(\CX_e^{sf})^{2n+1}(\CX_m^{sf})^2}{\CX_e^{sf}-1/\CX_e^{sf}}\right] ~,\\
\fb_{n.p.}^{(1)}&=\log[\CX_m^{n.p.(1)}]+\left(\frac{\CX_e^{sf}+1/\CX_e^{sf}}{\CX_e^{sf}-1/\CX_e^{sf}}\right)\log[\CX_e^{n.p.(1)}]\\
&\qquad\qquad+\log\left[1-\frac{(\CX_e^{sf})^{2n+1}(\CX_m^{sf})^2}{4(\CX_e^{sf}-1/\CX_e^{sf})^2}f_n(\CX_e^{sf})\right]~,
\end{split}\end{align}
Here $\fa_{n.p.}^{(1)},\fb_{n.p.}^{(1)}$ are the first of an infinite series of non-perturbative corrections to the semiclassical values of $\fa,\fb$
\begin{align}\begin{split}
\fa=\fa_{s.c.}+\sum_{i=1}^\infty \fa_{n.p.}^{(i)}\quad, \quad 
\fb=\fb_{s.c.}+\sum_{i=1}^\infty \fb_{n.p.}^{(i)}~,
\end{split}\end{align}
and $\CX_e^{n.p.(1)},\CX_m^{n.p.(1)}$ are the leading order non-perturbative corrections to the semi-flat expressions for $\CX_m,\CX_e$ which are given by  \cite{Gaiotto:2008cd}:
\begin{align}\begin{split}
&\CX_e^{n.p.(1)}(u,\Theta,\zeta)=\sum_{\gamma'=\gamma_n^\pm}\Omega(\gamma';u)\frac{\llangle \gamma',\half \alpha\rrangle}{4\pi i}e^{i \Theta\cdot \gamma'} \int_{\IR_+}\frac{d\zeta'}{\zeta'}\frac{\zeta'+\zeta e^{-i \alpha_{\gamma'}}}{\zeta'-\zeta e^{-i \alpha_{\gamma'}}}e^{-2\pi R|Z_{\gamma'}|(\zeta'+1/\zeta')}~,\\
&\CX_m^{n.p.(1)}(u,\Theta,\zeta)= \sum_{\gamma'=\gamma_n^\pm,\pm \alpha}\Omega(\gamma';u)\frac{\llangle \gamma',\half H_\alpha\rrangle}{4\pi i}e^{i \Theta\cdot \gamma'} \int_{\IR_+}\frac{d\zeta'}{\zeta'}\frac{\zeta'+\zeta e^{-i \alpha_{\gamma'}}}{\zeta'-\zeta e^{-i \alpha_{\gamma'}}}e^{-2\pi R|Z_{\gamma'}|(\zeta'+1/\zeta')}~,
\end{split}\end{align}
Using the BPS indices \eqref{SYMBPSInd}, these integrals simplify to 
\begin{align}\begin{split}
&\CX_e^{n.p.(1)}=\frac{i}{\pi } \sum_{n\in \IZ}\int_{\IR_+}\frac{d\zeta'}{\zeta'}\frac{ \sin(\theta_m+n \theta_e)(\zeta^{\prime 2}+\zeta^2)+2\zeta^\prime\zeta \cos(\theta_m+n \theta_e)\sin(\alpha_{\gamma_n^+})}{\zeta^{\prime 2}+\zeta^2-2\zeta\cos(\alpha_{\gamma_n^+})}e^{-2\pi R|Z_{\gamma'}|(\zeta'+1/\zeta')}~,\\
&\CX_m^{n.p.(1)}=\sum_{n\in \IZ}\frac{ n }{2\pi i }\int_{\IR_+}\frac{d\zeta'}{\zeta'}\frac{ \sin(\theta_m+n \theta_e)(\zeta^{\prime 2}+\zeta^2)+2\zeta^\prime\zeta \cos(\theta_m+n \theta_e)\sin(\alpha_{\gamma_n^+})}{\zeta^{\prime 2}+\zeta^2-2\zeta\cos(\alpha_{\gamma_n^+})}e^{-2\pi R|Z_{\gamma'}|(\zeta'+1/\zeta')}\\
&\quad+ \frac{2 i}{\pi}\int_{\IR_+}\frac{d\zeta'}{\zeta'}\frac{\sin(\theta_e)(\zeta^{\prime 2}+\zeta^2)+2\zeta^\prime \zeta \cos(\theta_e)\sin(\alpha_{\alpha})}{\zeta^{\prime 2}+\zeta^2-2\zeta\cos(\alpha_{\alpha})}e^{-2\pi R|Z_{\gamma'}|(\zeta'+1/\zeta')}~.
\end{split}\end{align}
where above we have used the notation where the integral over $\zeta'$ has been mapped to the integral over the positive reals by the phase rotation $e^{i \alpha_{\gamma'}}={\rm phase}(Z_{\gamma'})$   and $\llangle \gamma,\gamma'\rrangle$ is the DSZ pairing of charges. 




Note that the $\CY_\gamma$ are functions of $u,\theta_e,\theta_m,\zeta$ on the Hitchin moduli space. Because of the relation between the $\fa,\fb$ and the $\CX_e,\CX_m$ in \eqref{eq:DarbouxtoFN} and \eqref{FAtoSN}, we clearly see that the $\fa,\fb$ must also be functions of $u,\theta_e,\theta_m,\zeta$. The explicit dependence of $\fa,\fb$ on $\zeta$ can be seen first fixing a point in Hitchin moduli space 
with fixed coordinates $(u,\theta_e,\theta_m)$, 
and then studying the Fenchel-Nielsen coordinates as functions of the complex structure $\zeta$. 


\subsubsection{Index Theorem}

We can now use these coordinate transformations to determine an index formula as follows:
\begin{enumerate}
\item Calculate the localization computation for the expectation value of the given line operator:
\be
\langle L_{p,0}\rangle_{Loc}= \sum_{|\vv|<|P|}\frac{e^{(\vv,\fb)}}{\sinh^{|\vv|}( \fa)}\cdot \lim_{\xi\to 0}\int_{\widetilde\CM^{\xi}(P,\vv)} e^{\omega+\mu_T}\widehat{A}(T\widetilde\CM)~,
\ee
where $\widetilde\CM^{\xi}_{KN}(P,\vv)=\widetilde\CM^{\xi}_{KN}(\vk,\vw)$ is the corresponding Kronheimer-Nakajima quiver variety as described in Section \ref{sec:bubbling}. In this example it evaluates to
\be\label{LP0SYM}
\langle L_{p,0}\rangle_{Loc}=\left(\frac{e^\fb+e^{-\fb}}{2\sinh(\fa)}\right)^p~. 
\ee

\item Perform the change of coordinates:
\be
\fa\mapsto\log(f_n-\sqrt{f_n^2-1})\quad,\quad \fb\mapsto\log\left(\frac{\sqrt{a-\CX_e}}{\sqrt{a^{2n+1}(a \CX_e-1)}}\right)~,
\ee
in the localization result, where $f_n$ is given in equation \eqref{fnWilson} and $a=e^{\fa}$ as a function of $\CX_m,\CX_e$. 
\item Expand the $\langle L_{p,0}\rangle_{Loc}$ as a Laurent series in $\CX_m, \CX_e$:
\be
\langle L_{p,0}\rangle_{Loc}=\sum_{n_1,n_2}C_{n_m,n_e} \CX_m^{n_m}, \CX_e^{n_e}~.
\ee
\item Identify the coefficient of the $\CX_m^{n_m} \CX_e^{n_e}$ term, $C_{n_m,n_e}$, with the index:
\be
Ind\Big[\slashed{D}^{\CY}\Big]_{\fMM,c_n}^{\gamma_e}=C_{n_m,n_e}~,
\ee
where
\be
\zeta\in c_n\quad,\quad \fMM=\fMM(P,\gamma_m)\quad,\quad \gamma=\frac{n_e}{2}\alpha \oplus n_m H_\alpha\quad,\quad P=p H_\alpha~.
\ee
\end{enumerate}
%

After performing the Laurent expansion for  $\langle L_{p,0}\rangle $ given in \eqref{LP0SYM} in terms of the Darboux coordinates in the $c_n$ chamber, we have an expression
%
for the graded index of the twisted Dirac operator $\slashed{D}^{\CY}$ on singular monopole moduli space:
\begin{align}\begin{split}
{\rm Ind}&\Big[\slashed{D}^{\CY}\Big]^{\gamma_e=\frac{n_e}{2}\alpha}_{\fMM(P,\gamma_m;X_\infty)}=\sum_{m=0}^\infty\sum_{j=0}^p\sum_{k=0}^{2p} \sum_{\ell=0}^\infty \sum_{i=0}^\infty \sum_{q=0}^k\sum_{d_1=0}^{\substack{j+m+q+i\\+(2n+1)(p-k)}}\sum_{d_2=0}^{\lfloor d_1/2\rfloor}\sum_{d_3}^\infty
\left(\begin{array}{c}
p+m-1\\m
\end{array}\right)\left(\begin{array}{c}
2p\\k
\end{array}\right)\\&\times
\left(\begin{array}{c}
p+\ell-1\\\ell
\end{array}\right)
\left(\begin{array}{c}
k+i-1\\i
\end{array}\right)\left(\begin{array}{c}
p\\j
\end{array}\right)\left(\begin{array}{c}
k\\q
\end{array}\right)\left(\begin{array}{c}
j+m+q+i+(2n+1)(p-k)\\d_1
\end{array}\right)\left(\begin{array}{c}
\lfloor d_1/2 \rfloor\\d_2
\end{array}\right)\left(\begin{array}{c}
2d_3\\d_2
\end{array}\right)\\&\times
(-1)^{j+q+d_2}\frac{2^{-2p}}{(1-2d_3)}\times\begin{cases}
\sum_{i_1+i_2=2\ell-2d_2-2d_3}\left(\begin{array}{c}
2\ell-2d_2-2d_3\\i_1,i_2
\end{array}\right)(-1)^{i_1}\Delta_{n_e,n_m}&\ell-d_2-d_3>0\\
\sum_{i_1=0}^\infty \sum_{i_2=0}^{i_1}\left(\begin{array}{c}
i_1+2d_2+2d_3-2\ell-1\\i_1
\end{array}\right)\left(\begin{array}{c}
i_1\\i_2
\end{array}\right)(-1)^{i_1} \Delta_{n_e,n_m}&\ell-d_2-d_3<0
\end{cases}
\end{split}\end{align}
where $\Delta_{n_e,n_m}$ is a delta function that restricts the sum over the $\{m,j,k,\ell,i,q,d_i,i_i\}$ such that
\begin{align}\begin{split}
&n_m=\begin{cases}
i_2 & \ell-d_2-d_3>0\\
2(\ell-d_2-d_3)-i_1& \ell-d_2-d_3<0
\end{cases}~,\\
&n_e=\begin{cases}
2 i_1+(2n+2) i_2+2(d_2+d_3-\ell)+i +j+k-m-p & \ell-d_2-d_3>0\\
-(2n+2)i_1+2i_2-2(2n+1)(d_2+d_3-\ell)+i +j+m-p& \ell-d_2-d_3<0
\end{cases}~,
\end{split}\end{align}
are fixed. Additionally,
\be
P={\rm diag}(p,-p)\quad, \qquad \zeta\in c_n\quad,\qquad \gamma_m=n_mH_\alpha~.
\ee
This index formula for the case of $SU(2)$ SYM theory is also found in \cite{Moore:2015szp}.

\subsubsection{Characteristic Numbers}

Now by expressing $\CY_\gamma$ in terms of Fenchel-Nielsen coordinates, we can perform a Laurent expansion with respect to the exponential Fenchel-Nielsen coordinate $e^{ \fb}$. This will allow us to isolate the characteristic number. By using the equations for the Darboux coordinates in terms of Fenchel-Nielsen coordinates \eqref{eq:DarbouxtoFN}, we get the expansion
\begin{align}\begin{split}\label{MainCharNum}
\lim_{\substack{\xi\to0}}& \int_{\widetilde\CM_{KN}^\xi(P,\vv)} e^{\omega+\mu_T}\widehat{A}(T\widetilde{\CM}_{KN})=
\sum_{0\leq n_m,n_e\leq p}\Bigg\{\fro(n_m,n_e;c_n)Q_1^{(n_m,n_e)}(a;c_n)\\&\qquad
+\fro(-n_m,n_e;c_n)Q_2^{(n_m,n_e)}(a;c_n)
+\fro(-n_m,-n_e;c_n)Q_3^{(n_m,n_e)}(a;c_n)\\&
\qquad\qquad\qquad\qquad\hspace{4cm}
+\fro(n_m,-n_e;c_n)Q_4^{(n_m,n_e)}(a;c_n)\Bigg\}~.
\end{split}\end{align}
where
\begin{align}\begin{split}
&P={\rm diag}(p,-p)\quad, \quad \vv={\rm diag}(v,-v)~,\\&
\fro(n_m,n_e;c_n)=\fro(\gamma;c_n)\quad, \quad \gamma=n_m H_I\oplus n_e \half \alpha~,
\end{split}\end{align}
and the $Q_i^{(n_m,n_e)}(a,v;c_n)$ are different rational functions of $a$, defined as
\begin{align}\begin{split}
&Q_1^{(n_m,n_e)}(a,v;c_n)=\sum_{i_1=0}^{n_e} \sum_{j_1=0}^{\left[\substack{\frac{2n_m+v}{2}-i_1\\2nn_m}\right]}\sum^\prime_{\substack{i_2+j_2=\\\frac{v-n_m}{2}-i_1-j_1}}\left(\begin{array}{c}
n_e\\i_1
\end{array}\right)\left(\begin{array}{c}
i_2+n_e-1\\i_2
\end{array}\right)\left(\begin{array}{c}
2n_m\, n\\j_1
\end{array}\right)\\&\times
\left(\begin{array}{c}
j_2+2n_m(n+1)-1\\j_2
\end{array}\right)(-1)^{2n_m+i_2+j_2}a^{2n(i_1+j_2)+(2n+2)(i_2+j_1)+n_e-2(n_m-v)}(1-a^2)^{2(n_m+v)}
~,\\
&Q_2^{(n_m,n_e)}(a,v;c_n)=\sum_{i_1=0}^{n_e}\sum_{j_1=0}^{\left[\substack{\frac{2n_m+v}{2}-i_1\\2n_m(n+1)}\right]}\sum^\prime_{\substack{i_2+j_2=\\\frac{n_m+v}{2}-i_1-j_1}}\left(\begin{array}{c}
n_e\\i_1
\end{array}\right)\left(\begin{array}{c}
i_2-n_e-1\\i_2
\end{array}\right)\left(\begin{array}{c}
2n_m(n+1)\\j_1
\end{array}\right)\\&\quad\times\left(\begin{array}{c}
j_2+2n_m\,n-1\\j_2
\end{array}\right)
(-1)^{i_2+j_2+2n_m}\,a^{2n(i_1+j_1)+(2n+2)(i_2+j_2)+n_e+2n_m-2v}(1-a^2)^{2(v-n_m)}~,\\&
Q_3^{(n_m,n_e)}(a,v;c_n)=\sum_{i_1=0}^{n_e}\sum_{j_1=0}^{\left[\substack{\frac{2n_m+v}{2}-i_1\\2n_m(n+1)}\right]}\sum^\prime_{\substack{i_2+j_2=\\\frac{n_m+v}{2}-i_1-j_1}}\left(\begin{array}{c}
n_e\\i_1
\end{array}\right)\left(\begin{array}{c}
i_2-n_e-1\\i_2
\end{array}\right)\left(\begin{array}{c}
2n_m(n+1)\\j_1
\end{array}\right)\\&\quad\times\left(\begin{array}{c}
j_2+2n_mn-1\\j_2
\end{array}\right)
(-1)^{i_2+2n_m+j_2}\,a^{2n(i_1+j_1)+(2n+2)(i_1+j_2)+2n_m-n_e-2v}(1-a^2)^{2(v-n_m)}~,\\&
Q_4^{(n_m,n_e)}(a,v;c_n)=\sum_{i_1=0}^{n_e}\sum_{j_1=0}^{\left[\substack{\frac{2n_m+v}{2}-i_1\\2nn_m}\right]}\sum^\prime_{\substack{i_2+j_2=\\\frac{v-n_m}{2}-i_1-j_1}}\left(\begin{array}{c}
n_e\\i_1
\end{array}\right)\left(\begin{array}{c}
i_2+n_e-1\\i_2
\end{array}\right)\left(\begin{array}{c}
2n_m\,n\\j_1
\end{array}\right)\\&\times\left(\begin{array}{c}
j_2+2n_m(n+1)-1\\j_2
\end{array}\right)
(-1)^{i_2+j_2+2n_m}a^{2n(i_2+j_2)+(2n+2)(i_1+j_1)-2n_m-n_e-2v}(1-a^2)^{2(n_m+v)}
~.
\end{split}\end{align}
Here
we use the notation
\be
\sum_{i=0}^{\left[\substack{m\\n}\right]}= \sum_{i=0}^{{\rm min}[m,n]}\quad, \quad \left(\begin{array}{c}-1\\0\end{array}\right)=1\quad, \quad \sum^\prime_{i_2+j_2=...}=\begin{cases}
\sum_{i_2+j_2=...}&n_e,n_m\neq0\\
\sum_{\substack{j_2=...\\i_2=0}}&n_e=0~,~n_m\neq0\\
\sum_{\substack{i_2=...\\j_2=0}}&n_m=0~,~n_e\neq0\\
\sum_{i_2,j_2=0}&n_e=n_m=0~. 
\end{cases}
\ee
and the sums are restricted such that $\sum_{n=a}^b$ is identically zero for $b<a$.

Note that in both the formulas for the index of $\slashed{D}^\CY$ and the characteristic numbers on $\CM_{KN}$, there is a clear mixing of framed BPS states of magnetic charge $\gamma_m$ among many characteristic numbers for different. This suggests that there is a very non-trivial relationship between framed BPS states and the geometry of singular monopole moduli space since the $\CM_{KN}$ are transversal slices to singular strata in $\fMM(P,\gamma_m;X_\infty)$. It is an interesting challenge to differential geometers to try to prove such relations.

\subsubsection{Explicit Example: $\langle L_{2,0}\rangle$ in $SU(2)$ SYM}

We can illustrate the above formulas for the index of $\slashed{D}^{\CY}$ and the characteristic numbers on $\CM_{KN}(P,\vv)$  with the non-trivial example of the next-to-minimal 't Hooft defect: $L_{2,0}$. 

Let us first demonstrate  the index theorem by calculating the index of $\slashed{D}^\CY$. In our example, the expectation value from localization can be written 
\be
\langle L_{2,0}\rangle_{Loc}=\frac{2\cosh(2\fb)}{\sinh^2(\fa)}+Z_{mono}(\fa;2,0)~,
\ee
where 
\be
Z_{mono}(\fa;2,0)=\lim_{\xi\to0}\int_{\widetilde\CM^\xi(2,0)} e^{\omega+\mu_T}\widehat{A}(T\widetilde{M})~,
\ee
is the characteristic number on the Kronheimer-Nakajima space defined by the quiver 

\begin{center}
\begin{tikzpicture}[
cnode/.style={circle,draw,thick,minimum size=9mm},snode/.style={rectangle,draw,thick,minimum size=9mm}]
\node[cnode] (1) {1};
\node[snode] (2) [below=.5cm  of 1]{2};
\draw[-] (1) -- (2);
\end{tikzpicture}
\end{center}
as described in Section \ref{sec:sec2}. 

For this example, the characteristic number evaluates to \cite{Brennan:2018yuj}
\be
Z_{mono}(\fa;2,0)=\frac{2}{4\sinh^2(\fa)}~. 
\ee
Let us compute the index theorem for the chamber $c_1$. In this chamber, the coordinate transformation is of the form 
\begin{align}\begin{split}
&a=e^\fa=f_1-\sqrt{f_1^2-1}\quad, \quad f_1=\half \CX_e+\frac{1}{2\CX_e}+\frac{\CX_m^2\CX_e^{2n+1}}{2}~,\\
&b=e^\fb=\sqrt{\frac{f_1-\sqrt{f_1^2-1}-\CX_e}{(f_1-\sqrt{f_1^2-1})^4\CX_e-(f_1-\sqrt{f_1^2-1})^3}}~. 
\end{split}\end{align}
Plugging this into the full expectation value
\be
\langle L_{2,0}\rangle_{Loc}=\frac{2\cosh(2\fb)}{\sinh^2(\fa)}+\frac{1}{2\sinh^2(\fa)}~,
\ee 
yields the Darboux expansion 
\be
\langle L_{2,0}\rangle_{Loc}\Big{|}_{\fa,\fb\mapsto \CX_m,\CX_e}=\frac{1}{\CX_m^2}+\CX_e^4\CX_m^2+2\CX_e^2~,
\ee
in terms of the spectral network coordinates in the $c_1$ chamber. 
Note that this matches the direct computation from spectral networks \eqref{SYMLineDef} \cite{Moore:2015szp}. 

From this expansion we can read off the indices of the Dirac operator:
\be
{\rm Ind}\Big[\slashed{D}^{\CY}\Big]^{\gamma_e=\frac{n_e}{2}\alpha}_{\fMM(P,\gamma_m;X_\infty)}=\begin{cases}
1&\gamma=H_\alpha\oplus 2\alpha\\
1&\gamma=-H_\alpha\\
2&\gamma=\alpha\\
0&else
\end{cases}
\ee
where $P=\half{\rm diag}(2,-2)$ and $\zeta\in c_1$.

Now let us perform the inverse coordinate substitution to derive the characteristic number from the spectral network computation. Let us start with the expectation value of $L_{2,0}$ from the spectral network associated with $\zeta\in c_1$:
\be\label{spectnetL20}
\langle L_{2,0}\rangle_{\zeta\in c_1}=\frac{1}{\CX_m^2}+\CX_e^4\CX_m^2+2\CX_e^2~. 
\ee
The coordiante transformation \eqref{eq:DarbouxtoFN} now takes the form 
\be
\CX_m=-\frac{b(a^2-1)(1+a^4 b^2)}{a(1+a^2 b^2)^2}\quad, \quad \CX_e=\frac{a(1+a^2b^2)}{(1+a^4 b^2)}~. 
\ee
Plugging this into \eqref{spectnetL20} we find 
\be
\langle L_{2,0}\rangle_{\zeta\in c_1}\Big{|}_{\CX_m,\CX_e\mapsto \fa,\fb}=\frac{(b+1/b)^2}{(a-1/a)^2}~,
\ee
which indeed matches with the localization computation. Expanding this in powers of $b=e^\fb$, we see that the $0^{th}$ order term ($Z_{mono}(\fa;2,0)$) is given by 
\be
Z_{mono}(\fa;2,0)=\frac{2}{(a-1/a)^2}=\frac{2}{4\sinh^2(\fa)}~. 
\ee

We can also derive this result from the full formula for the characteristc number. Using the data
\be\label{fros}
\fro(-1,0;c_1)=\fro(1,4;c_1)=1\quad, \quad \fro(0,2;c_1)=2~,
\ee
 the characteristic number formula \eqref{MainCharNum} reduces to 
\begin{align}\begin{split}
\lim_{\substack{\xi\to0}}& \int_{\widetilde\CM_{KN}^\xi(2,0)} e^{\omega+\mu_T}\widehat{A}(T\widetilde{\CM}_{KN})=\\
&
\fro(1,4;c_1)Q_1^{(1,4)}(a;c_1)
+\fro(0,2;c_1)Q_1^{(0,2)}(a;c_1)
+\fro(-1,0;c_1)Q_2^{(1,0)}(a;c_1)~.
\end{split}\end{align}
Evaluating the polynomials, we find 
\begin{align}\begin{split}\label{polies}
&Q_1^{(1,4)}(a;c_1)=0~,\\
&Q_1^{(0,2)}(a;c_1)=\sum_{i_1=0}^2\sum_{j_1=0}^{-i_1}\sum_{i_2=0}\left(\begin{array}{c}2\\i_1\end{array}\right)\left(\begin{array}{c}i_2+1\\i_2\end{array}\right)(-1)^{i_2+j_2}a^{2(i_1+j_2)+(4(i_2+j_1)+2}=a^2~,\\
&Q_2^{(1,0)}(a;c_1)=\sum_{j_1=0}^1\sum_{j_2=1-j_1}\left(\begin{array}{c}4\\j_1\end{array}\right)\left(\begin{array}{c}j_2+1\\j_2\end{array}\right)(-1)^{j_2}\frac{a^{2j_1+4j_2+2}}{(1-a^2)^2}=\frac{-2 a^6+4 a^4}{(1-a^2)^2}~. 
\end{split}\end{align}
Combining these results with the framed BPS indices \eqref{fros}, the full formula for the characteristic number evaluates to 
\begin{align}\begin{split}
\lim_{\substack{\xi\to0}}& \int_{\widetilde\CM_{KN}^\xi(2,0)} e^{\omega+\mu_T}\widehat{A}(T\widetilde{\CM}_{KN})=
2a^2+
\frac{-2 a^6+4 a^4}{(1-a^2)^2}=\frac{2}{(a-1/a)^2}~,
\end{split}\end{align}
matching the result from direct computation.

\subsection{Comments on the $\CN=2^\ast$ Theory}

Here we would like to make some clarifying comments on the $SU(2)$ $\CN=2^\ast$ theory and the Fenchel-Nielsen locus in this theory. 
In the case of the $\CN=2^\ast$ theory the UV curve $C$ is given by the once punctured torus. The algebra of line operators of this theory can be generated by the three simple line operators $L_{\gamma_{(1,0)}},L_{\gamma_{(0,1)}}$, and $L_{\gamma_{(1,1)}}$. Note that there are three generating operators because the homology lattice is generated by a cycles that wrap the $A$-cycle, $B$-cycle, and the puncture. 

\begin{figure}
\begin{center}
\includegraphics[scale=0.7]
{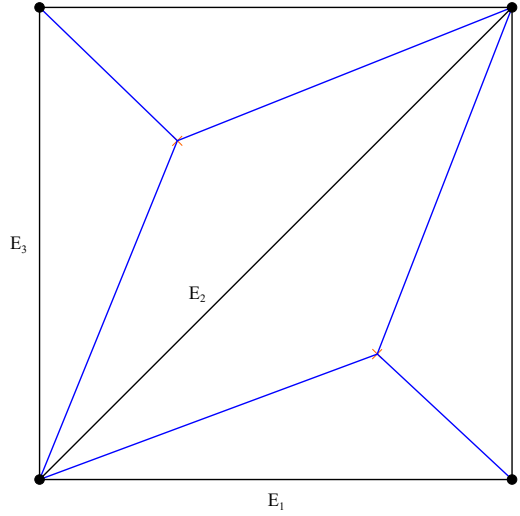}
\end{center}
\caption{This figure shows a generic WKB spectral network (blue) on the punctured torus. This corresponds to the triangulation given by the (black) edges $E_{1,2,3}$ where the rectangle is periodically identified and the puncture is located at the identified corners. }
\label{fig:squaretorus}
\end{figure}

A generic spectral network associated to the $SU(2)$ $\CN=2^\ast$ theory is given by an ideal triangulation of $C$ as in Figure \ref{fig:squaretorus}. In each chamber of the $\zeta$-plane $c$, the 
charge lattice is spanned by three simple elements $\gamma_i[c]$ for $i=1,2,3$ such that 
\be
\langle \gamma_i[c],\gamma_{i+1}[c]\rangle=2\quad, \quad \gamma_1[c]+\gamma_2[c]+\gamma_3[c]=\gamma_f~. 
\ee
Given a particular choice of chamber $c_0$ we can identify 
\be
\gamma_1[c_0]=-\alpha\oplus \gamma_f\quad, \quad \gamma_2[c_0]=-H_\alpha\quad, \quad \gamma_3[c_0]=H_\alpha\oplus \alpha~. 
\ee
In such a chamber, the expectation values of the line operators can be expanded in terms as 
\begin{align}\begin{split}
\langle L_{\gamma_{(1,0)}}\rangle=\sqrt{\CY_{\gamma_2}\CY_{\gamma_3}}+\frac{1}{\sqrt{\CY_{\gamma_2} \CY_{\gamma_3}}}+\sqrt{\frac{\CY_{\gamma_3}}{\CY_{\gamma_2}}}~,\\
\langle L_{\gamma_{(0,1)}}\rangle=\sqrt{\CY_{\gamma_3}\CY_{\gamma_1}}+\frac{1}{\sqrt{\CY_{\gamma_3}\CY_{\gamma_1}}}+\sqrt{\frac{\CY_{\gamma_1}}{\CY_{\gamma_3}}}~,\\
\langle L_{\gamma_{(1,1)}}\rangle=\sqrt{\CY_{\gamma_2}\CY_{\gamma_1}}+\frac{1}{\sqrt{\CY_{\gamma_2}\CY_{\gamma_1}}}+\sqrt{\frac{\CY_{\gamma_2}}{\CY_{\gamma_1}}}~. 
\end{split}\end{align}
Here $\CY_{\gamma_i}$ is the spectral network coordinate corresponding to the edge $E_i$.

\begin{figure}
\begin{center}
\includegraphics[scale=0.7,clip,trim=0cm 0cm 0cm 3cm]
{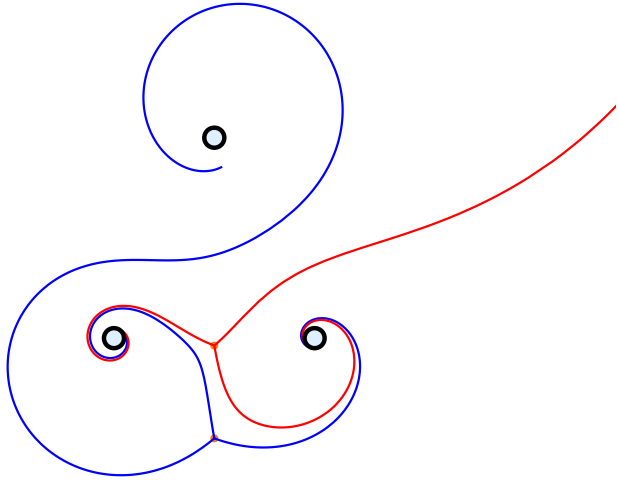}
\end{center}
\caption{This figure shows a generic WKB spectral network on the punctured torus. The punctured torus is presented as a trinion with two boundary circles identified. These are the two lower circles in the above figure. We choose the A-cycle to be defined by the boundary of these circles. 
Here the lines are the walls of the corresponding WKB spectral network. While it is not drawn here due to computational limitations, the walls corresponding to the open paths run to the (upper) puncture. }
\label{fig:roundtorus}
\end{figure}

In the $SU(2)$ $\CN=2^\ast$ theory, the Fenchel-Nielsen locus is  defined by 
\be
\bar{m} \int_A \lambda_{SW}\in \IR\quad, \quad \int_A \zeta^{-1}\lambda_{SW}\in \IR~,
\ee
where $\zeta$ is the phase defining the line operator (and corresponding WKB spectral network). As it turns out, this coincides with the exceptional locus 
\be
\CE=\bigcup_i \CE_i\quad, \quad \CE_i=\left\{u\in \CB~|~Z(\gamma_i;u)/m>0~,~{\rm Arg}[Z(\gamma_{i+1};u)]<{\rm Arg}[Z(\gamma_{i-1};u)]\right\}~,
\ee
from \cite{LonghiThesis}. Here mathematical simplifications arise that allow for the exact computation of the spectrum generator which encodes the entire spectrum of BPS states. 

\begin{figure}
\begin{center}
\includegraphics[scale=0.8,clip,trim=0cm 0cm 0cm 4cm]
{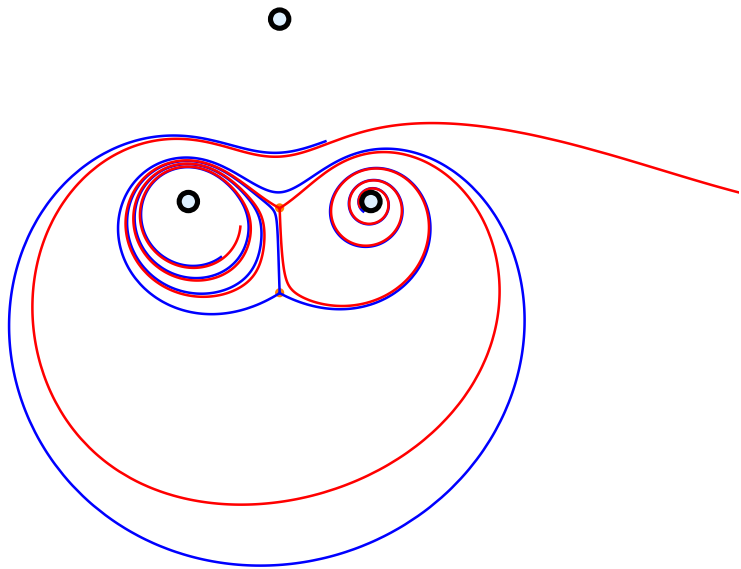}
\end{center}
\caption{This figure shows the behavior of the generic WKB spectral network on the punctured torus from Figure \ref{fig:roundtorus} as it approaches the Fenchel-Nielsen spectral network (left in Figure \ref{fig:mol}). Again, while it is not drawn here due to computational limitations, the walls corresponding to the open paths run to the (upper) puncture. 
}
\label{fig:FNlimit}
\end{figure}

As we approach to the Fenchel-Nielsen locus, we cross an infinite number of $\CK$-walls in passing through the chambers $c_n$ with increasing $n$. Mathematically, crossing the $\CK$-wall going from chamber $c_n\to c_{n+1}$ corresponds to mutating along one of basis elements of the charge lattice in $c_n$, $\gamma_i[c_n]$. 
As discussed in \cite{Gaiotto:2010be}, this transformation keeps the 
three-term expansion of the $\langle L_\gamma\rangle$ that have explicit 
$\CY_{\gamma_i}$ dependence  but increases the complexity of the 
$\langle L_\gamma\rangle$ that are independent of $\CY_{\gamma_i}$. This 
leads to a fairly simple change of variables between
the complexified Fenchel-Nielsen coordinates and the $\CY_{\gamma_i}$ 
given by \cite{Dimofte:2011jd}
\begin{align}\begin{split}
&\sqrt{\CY_{\gamma_1}} = \frac{i}{\ell} \frac{\tilde\beta - \tilde\beta^{-1}}{\tilde\beta \lambda - (\tilde\beta\lambda)^{-1}}\quad, \quad 
\sqrt\CY_{\gamma_2} = i \frac{\tilde\beta \lambda - (\tilde\beta\lambda)^{-1} 
}{\lambda-\lambda^{-1}}\quad, \quad 
\sqrt\CY_{\gamma_3} =- i \frac{\lambda-\lambda^{-1}}{\tilde\beta - \tilde\beta^{-1}}~,\\
&\lambda=e^\fa\quad, \quad \ell=e^{ m}\quad, \quad \tilde\beta=e^{\fb}\sqrt{\frac{e^{\fa+m}-e^{-\fa-m}}{e^{\fa-m}-e^{-\fa+m}}}~.
\end{split}\end{align}
Sending $\zeta\to \zeta_{FN}$ acts on the corresponding spectral network as in Figure \ref{fig:FNlimit}. This makes it obvious that the Wilson line $\langle L_{\gamma_{(1,0)}}\rangle$, which is the holonomy around one of the resolved punctures, keeps a three term expansion. And further, from the properties of a Fenchel-Nielsen spectral network, we see that the expression for $\langle L_{\gamma_{(1,0)}}\rangle$ becomes a two term expansion in the limit $\zeta\to \zeta_{FN}$. We believe mirrors the same behavior of the expectation value of the Wilson line in the $SU(2)$ N$_f=0$ theory as discussed in the previous section.

\section*{Acknowledgements}

TDB and GWM would like to thank Anindya Dey and Andy Neitzke for fruitful collaborations on related material and Tudor Dimofte, Davide Gaiotto, and Feng Luo for enlightening discussions. TDB and GWM would like to extend a special thanks to Pietro Longhi for extensive comments on the draft, enlightening discussion, and assistance with figures. TDB and GWM are supported by DOE grant DOE-SC0010008.


\begin{thebibliography}{99}






\bibitem{Alday:2009fs}
  L.~F.~Alday, D.~Gaiotto, S.~Gukov, Y.~Tachikawa and H.~Verlinde,
  ``Loop and surface operators in N=2 gauge theory and Liouville modular geometry,''
  JHEP {\bf 1001}, 113 (2010)
  doi:10.1007/JHEP01(2010)113
  [arXiv:0909.0945 [hep-th]].

\bibitem{Assel:2019iae} 
  B.~Assel and A.~Sciarappa,
  ``On Monopole Bubbling Contributions to 't Hooft Loops,''
  arXiv:1903.00376 [hep-th].

\bibitem{Brennan:2016znk}
  T.~D.~Brennan and G.~W.~Moore,
  ``A note on the semiclassical formulation of BPS states in four-dimensional $N=$ 2 theories,''
  PTEP {\bf 2016}, no. 12, 12C110 (2016)
  doi:10.1093/ptep/ptw159
  [arXiv:1610.00697 [hep-th]].

\bibitem{Brennan:2018yuj}
  T.~D.~Brennan, A.~Dey and G.~W.~Moore,
  ``On 't Hooft Defects, Monopole Bubbling and Supersymmetric Quantum Mechanics,''
  arXiv:1801.01986 [hep-th].

\bibitem{Brennan:2018moe}
  T.~D.~Brennan,
  ``Monopole Bubbling via String Theory,''
  arXiv:1806.00024 [hep-th].

\bibitem{Brennan:2018rcn}
  T.~D.~Brennan, A.~Dey and G.~W.~Moore,
  ``'t Hooft Defects and Wall Crossing in SQM,''
  arXiv:1810.07191 [hep-th].





\bibitem{Dehn}
M.~Dehn,
``Lecture Notes from Breslau,'' 1992. The Archives of the University of Texas at Austin.


\bibitem{Diaconescu:1996rk}
  D.~E.~Diaconescu,
  ``D-branes, monopoles and Nahm equations,''
  Nucl.\ Phys.\ B {\bf 503}, 220 (1997)
  doi:10.1016/S0550-3213(97)00438-0
  [hep-th/9608163].

\bibitem{Dimofte:2011jd}
  T.~Dimofte and S.~Gukov,
  ``Chern-Simons Theory and S-duality,''
  JHEP {\bf 1305}, 109 (2013)
  doi:10.1007/JHEP05(2013)109
  [arXiv:1106.4550 [hep-th]].


\bibitem{Drukker:2009id}
  N.~Drukker, J.~Gomis, T.~Okuda and J.~Teschner,
  ``Gauge Theory Loop Operators and Liouville Theory,''
  JHEP {\bf 1002}, 057 (2010)
  doi:10.1007/JHEP02(2010)057
  [arXiv:0909.1105 [hep-th]].
  
\bibitem{Drukker:2009tz} 
  N.~Drukker, D.~R.~Morrison and T.~Okuda,
  ``Loop operators and S-duality from curves on Riemann surfaces,''
  JHEP {\bf 0909}, 031 (2009)
  doi:10.1088/1126-6708/2009/09/031
  [arXiv:0907.2593 [hep-th]].

\bibitem{FLuo}
F. Luo, \textit{private communication.}

%






\bibitem{Gaiotto:2009we}
  D.~Gaiotto,
  ``N=2 dualities,''
  JHEP {\bf 1208}, 034 (2012)
  doi:10.1007/JHEP08(2012)034
  [arXiv:0904.2715 [hep-th]].

\bibitem{Gaiotto:2008sa}
  D.~Gaiotto and E.~Witten,
  ``Supersymmetric Boundary Conditions in N=4 Super Yang-Mills Theory,''
  J.\ Statist.\ Phys.\  {\bf 135}, 789 (2009)
  doi:10.1007/s10955-009-9687-3
  [arXiv:0804.2902 [hep-th]].

\bibitem{Gaiotto:2008cd}
  D.~Gaiotto, G.~W.~Moore and A.~Neitzke,
  ``Four-dimensional wall-crossing via three-dimensional field theory,''
  Commun.\ Math.\ Phys.\  {\bf 299}, 163 (2010)
  doi:10.1007/s00220-010-1071-2
  [arXiv:0807.4723 [hep-th]].

\bibitem{Gaiotto:2009hg}
  D.~Gaiotto, G.~W.~Moore and A.~Neitzke,
  ``Wall-crossing, Hitchin Systems, and the WKB Approximation,''
  arXiv:0907.3987 [hep-th].

\bibitem{Gaiotto:2010be}
  D.~Gaiotto, G.~W.~Moore and A.~Neitzke,
  ``Framed BPS States,''
  Adv.\ Theor.\ Math.\ Phys.\  {\bf 17}, no. 2, 241 (2013)
  doi:10.4310/ATMP.2013.v17.n2.a1
  [arXiv:1006.0146 [hep-th]].

\bibitem{Gaiotto:2012rg}
  D.~Gaiotto, G.~W.~Moore and A.~Neitzke,
  ``Spectral networks,''
  Annales Henri Poincare {\bf 14}, 1643 (2013)
  doi:10.1007/s00023-013-0239-7
  [arXiv:1204.4824 [hep-th]].
\bibitem{Gaiotto:2012db}
  D.~Gaiotto, G.~W.~Moore and A.~Neitzke,
 ``Spectral Networks and Snakes,''
  Annales Henri Poincare {\bf 15}, 61 (2014)
  doi:10.1007/s00023-013-0238-8
  [arXiv:1209.0866 [hep-th]].

%


%
\bibitem{Gauntlett:1999vc}
  J.~P.~Gauntlett, N.~Kim, J.~Park and P.~Yi,
  ``Monopole dynamics and BPS dyons N=2 superYang-Mills theories,''
  Phys.\ Rev.\ D {\bf 61}, 125012 (2000)
  doi:10.1103/PhysRevD.61.125012
  [hep-th/9912082].

\bibitem{Gauntlett:2000ks}
  J.~P.~Gauntlett, C.~j.~Kim, K.~M.~Lee and P.~Yi,
  ``General low-energy dynamics of supersymmetric monopoles,''
  Phys.\ Rev.\ D {\bf 63}, 065020 (2001)
  doi:10.1103/PhysRevD.63.065020
  [hep-th/0008031].


\bibitem{Gomis:2011pf}
  J.~Gomis, T.~Okuda and V.~Pestun,
  ``Exact Results for 't Hooft Loops in Gauge Theories on $S^4$,''
  JHEP {\bf 1205}, 141 (2012)
  doi:10.1007/JHEP05(2012)141
  [arXiv:1105.2568 [hep-th]].

\bibitem{Hanany:1996ie}
  A.~Hanany and E.~Witten,
  ``Type IIB superstrings, BPS monopoles, and three-dimensional gauge dynamics,''
  Nucl.\ Phys.\ B {\bf 492}, 152 (1997)
  doi:10.1016/S0550-3213(97)00157-0, 10.1016/S0550-3213(97)80030-2
  [hep-th/9611230].

\bibitem{Hollands:2013qza}
  L.~Hollands and A.~Neitzke,
  ``Spectral Networks and Fenchel–Nielsen Coordinates,''
  Lett.\ Math.\ Phys.\  {\bf 106}, no. 6, 811 (2016)
  doi:10.1007/s11005-016-0842-x
  [arXiv:1312.2979 [math.GT]].


\bibitem{Ito:2011ea}
  Y.~Ito, T.~Okuda and M.~Taki,
  ``Line operators on $S^1\times R^3$ and quantization of the Hitchin moduli space,''
  JHEP {\bf 1204}, 010 (2012)
  Erratum: [JHEP {\bf 1603}, 085 (2016)]
  doi:10.1007/JHEP03(2016)085, 10.1007/JHEP04(2012)010
  [arXiv:1111.4221 [hep-th]].
%

\bibitem{Jeong:2018qpc} 
  S.~Jeong and N.~Nekrasov,
  ``Opers, surface defects, and Yang-Yang functional,''
  arXiv:1806.08270 [hep-th].

%
\bibitem{Kapustin:2006pk}
  A.~Kapustin and E.~Witten,
  ``Electric-Magnetic Duality And The Geometric Langlands Program,''
  Commun.\ Num.\ Theor.\ Phys.\  {\bf 1}, 1 (2007)
  doi:10.4310/CNTP.2007.v1.n1.a1
  [hep-th/0604151].

\bibitem{Klemm:1996bj}
  A.~Klemm, W.~Lerche, P.~Mayr, C.~Vafa and N.~P.~Warner,
  ``Selfdual strings and N=2 supersymmetric field theory,''
  Nucl.\ Phys.\ B {\bf 477}, 746 (1996)
  doi:10.1016/0550-3213(96)00353-7
  [hep-th/9604034].

%
%
\bibitem{Kronheimer:1990}
	P.~B.~ Kronheimer and H.~ Nakajima,
	"Yang-Mills Instantons on ALE Gravitional Instantons,"
	Math.\ Ann. \ (1990) \ {\bf 288}: 263. 

\bibitem{Liu0}
	J.~Liu,
	``On the Existence of Jenkins-Strebel Differentials,''
	Bull. of Lond. Math. Soc. {\bf 36}, 03 (2004).

\bibitem{Liu}
	J.~ Liu,
	``Jenkins-Strebel Differentials with Poles,"
	Comment.\ Math.\ Helv.\ {\bf 83}, 01 (2008).

\bibitem{LonghiThesis}
P.~Longhi, 
``The Structure of BPS Spectra,''
Ph.D. Thesis, Rutgers University Library (2015). 

\bibitem{Longhi:2016rjt}
  P.~Longhi and C.~Y.~Park,
  ``ADE Spectral Networks,''
  JHEP {\bf 1608}, 087 (2016)
  doi:10.1007/JHEP08(2016)087
  [arXiv:1601.02633 [hep-th]].

\bibitem{Longhi:2016bte}
  P.~Longhi and C.~Y.~Park,
  ``ADE Spectral Networks and Decoupling Limits of Surface Defects,''
  JHEP {\bf 1702}, 011 (2017)
  doi:10.1007/JHEP02(2017)011
  [arXiv:1611.09409 [hep-th]].


\bibitem{Martens:2006hu}
  J.~Martens,
  ``Equivariant volumes of non-compact quotients and instanton counting,''
  Commun.\ Math.\ Phys.\  {\bf 281}, 827 (2008)
  doi:10.1007/s00220-008-0501-x
  [math/0609841 [math-sg]].

\bibitem{Manton:1981mp}
  N.~S.~Manton,
  ``A Remark on the Scattering of BPS Monopoles,''
  Phys.\ Lett.\  {\bf 110B}, 54 (1982).
  doi:10.1016/0370-2693(82)90950-9

\bibitem{Manton:1993aa}
  N.~S.~Manton and B.~J.~Schroers,
  ``Bundles over moduli spaces and the quantization of BPS monopoles,''
  Annals Phys.\  {\bf 225}, 290 (1993).
  doi:10.1006/aphy.1993.1060

\bibitem{Moore:1997dj}
  G.~W.~Moore, N.~Nekrasov and S.~Shatashvili,
  ``Integrating over Higgs branches,''
  Commun.\ Math.\ Phys.\  {\bf 209}, 97 (2000)
  doi:10.1007/PL00005525
  [hep-th/9712241].

\bibitem{Moore:1998et}
  G.~W.~Moore, N.~Nekrasov and S.~Shatashvili,
  ``D particle bound states and generalized instantons,''
  Commun.\ Math.\ Phys.\  {\bf 209}, 77 (2000)
  doi:10.1007/s002200050016
  [hep-th/9803265].



\bibitem{Moore:2015szp}
  G.~W.~Moore, A.~B.~Royston and D.~Van den Bleeken,
  ``Semiclassical framed BPS states,''
  JHEP {\bf 1607}, 071 (2016)
  doi:10.1007/JHEP07(2016)071
  [arXiv:1512.08924 [hep-th]].
%
\bibitem{Moore:2015qyu}
  G.~W.~Moore, A.~B.~Royston and D.~Van den Bleeken,
  ``L$^2$-Kernels Of Dirac-Type Operators On Monopole Moduli Spaces,''
  arXiv:1512.08923 [hep-th].
%

%
%




%
\bibitem{Nakajima:2016guo}
  H.~Nakajima and Y.~Takayama,
  ``Cherkis bow varieties and Coulomb branches of quiver gauge theories of affine type $A$,''
  arXiv:1606.02002 [math.RT].
%
\bibitem{Nekrasov:2002qd}
  N.~A.~Nekrasov,
  ``Seiberg-Witten prepotential from instanton counting,''
  Adv.\ Theor.\ Math.\ Phys.\  {\bf 7}, no. 5, 831 (2003)
  doi:10.4310/ATMP.2003.v7.n5.a4
  [hep-th/0206161].
%

\bibitem{Nekrasov:2011bc} 
  N.~Nekrasov, A.~Rosly and S.~Shatashvili,
  ``Darboux coordinates, Yang-Yang functional, and gauge theory,''
  Nucl.\ Phys.\ Proc.\ Suppl.\  {\bf 216}, 69 (2011)
  doi:10.1016/j.nuclphysbps.2011.04.150
  [arXiv:1103.3919 [hep-th]].

\bibitem{PratoWu}
E.~Prato and S.~ Wu,
``Duistermaat-Heckman measures in a non-compact setting,''
 arXiv:alg-geom/9307005.

\bibitem{Thurston}
W.P.~Thurston,
``On the geometry and dynamics of diffeomorphisms of surfaces,''
Bull. Amer. Math. Soc. (N.S.) {\bf 19} (1988), no. 2, 417-431.

\bibitem{Tong:2014yla}
  D.~Tong and K.~Wong,
  ``Monopoles and Wilson Lines,''
  JHEP {\bf 1406}, 048 (2014)
  doi:10.1007/JHEP06(2014)048
  [arXiv:1401.6167 [hep-th]].
%
%
\bibitem{Witten:1997sc}
  E.~Witten,
  ``Solutions of four-dimensional field theories via M theory,''
  Nucl.\ Phys.\ B {\bf 500}, 3 (1997)
  doi:10.1016/S0550-3213(97)00416-1
  [hep-th/9703166].
%
%
%

\end{thebibliography}
\end{document}